\documentclass[fleqn,usenatbib]{mnras}

\usepackage{newtxtext,newtxmath}

\usepackage[T1]{fontenc}
\usepackage{ae,aecompl}

\usepackage{graphicx}	
\usepackage{amsmath}	
\usepackage{soul}

\title[GOTO prototype performance]{The Gravitational-wave Optical Transient Observer (GOTO): prototype performance and prospects for transient science}

\author[D. Steeghs et al.]{D. Steeghs \href{https://orcid.org/0000-0003-0771-4746}{\includegraphics[scale=0.05]{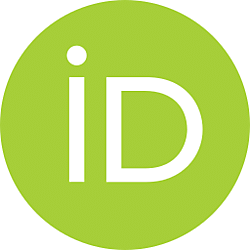}},$^{1,13}$\thanks{E-mail: D.T.H.Steeghs@warwick.ac.uk}
D. K. Galloway,$^{2,13}$ K. Ackley \href{https://orcid.org/0000-0002-8648-0767}{\includegraphics[scale=0.05]{ID.png}},$^{1,2,13}$ 
M. J. Dyer \href{https://orcid.org/0000-0003-3665-5482}{\includegraphics[scale=0.05]{ID.png}},$^{3}$
J. Lyman,$^{1}$ K. Ulaczyk,$^{1}$ 
\newauthor 
R. Cutter \href{https://orcid.org/0000-0001-8945-5551}{\includegraphics[scale=0.05]{ID.png}},$^{1}$
Y.-L. Mong,$^{2,13}$ V. Dhillon,$^{3,9}$ P. O'Brien,$^{4}$ G. Ramsay,$^{5}$ S.
Poshyachinda,$^{6}$ 
\newauthor 
R. Kotak,$^{7}$ L. K. Nuttall,$^{8}$ E.
Pall\'e,$^{9,15}$ R. P. Breton \href{https://orcid.org/0000-0001-8522-4983}{\includegraphics[scale=0.05]{ID.png}},$^{10}$ D. Pollacco,$^{1}$ E. Thrane,$^{2}$ 
\newauthor 
S. Aukkaravittayapun,$^{6}$ S. Awiphan,$^{6}$ U.
Burhanudin,$^{3}$ P. Chote,$^{1}$ A. Chrimes,$^{1}$ 
\newauthor 
E. Daw,$^{3}$ C. Duffy,$^{1,5}$ R. Eyles-Ferris,$^{4}$ B. Gompertz,$^{1}$ T.
Heikkil\"a,$^{7}$ P. Irawati,$^{6}$ 
\newauthor 
M. R. Kennedy \href{https://orcid.org/0000-0001-6894-6044}{\includegraphics[scale=0.05]{ID.png}},$^{10}$ T. Killestein,$^{1}$ H. Kuncarayakti \href{https://orcid.org/0000-0002-1132-1366}{\includegraphics[scale=0.05]{ID.png}},$^{11,14}$ A. J. Levan,$^{1,15}$ S. Littlefair,$^{3}$ 
\newauthor 
L. Makrygianni,$^{3}$ T. Marsh,$^{1}$ D. Mata-Sanchez \href{https://orcid.org/0000-0003-0245-9424}{\includegraphics[scale=0.05]{ID.png}},$^{9,10,16}$ S.
Mattila,$^{7}$ J. Maund,$^{3}$ J. McCormac,$^{1}$ 
\newauthor 
D. Mkrtichian,$^{6}$J. Mullaney,$^{3}$ K. Noysena \href{https://orcid.org/0000-0001-9109-8311}{\includegraphics[scale=0.05]{ID.png}},$^{6}$ M. Patel,$^{4}$ E. Rol,$^{2}$ U. Sawangwit,$^{6}$ E. R. Stanway,$^{1}$
\newauthor 
R. Starling,$^{4}$ P. Str\o{}m\href{https://orcid.org/0000-0002-7823-1090}{\includegraphics[scale=0.05]{ID.png}},$^{1}$ S. Tooke,$^{4}$ R. West,$^{1}$ D. J. White,$^{12}$ K. Wiersema$^{1}$
\\
$^{1}$Department of Physics, University of Warwick, Gibbet Hill Road, Coventry CV4 7AL, UK \\
$^{2}$School of Physics \& Astronomy, Monash University, Clayton VIC 3800, Australia \\
$^{3}$Department of Physics and Astronomy, University of Sheffield, Sheffield S3 7RH, UK \\
$^{4}$School of Physics \& Astronomy, University of Leicester, University Road, Leicester LE1 7RH, UK \\
$^{5}$Armagh Observatory \& Planetarium, College Hill, Armagh, BT61 9DG \\
$^{6}$National Astronomical Research Institute of Thailand, 260 Moo 4, T. Donkaew, A. Maerim, Chiangmai, 50180 Thailand \\
$^{7}$Department of Physics \& Astronomy, University of Turku, Vesilinnantie 5, Turku, FI-20014, Finland \\
$^{8}$University of Portsmouth, Portsmouth, PO1 3FX, UK \\
$^{9}$Instituto de Astrof\'isica de Canarias, E-38205 La Laguna, Tenerife, Spain \\
$^{10}$Jodrell Bank Centre for Astrophysics, Department of Physics and Astronomy, The University of Manchester, Manchester M13 9PL, UK \\
$^{11}$ Tuorla Observatory, Department of Physics and Astronomy, FI-20014 University of Turku, Finland \\
$^{12}$EPCC, University of Edinburgh, Bayes Centre, 47 Potterrow, Edinburgh EH8 9BT, UK \\
$^{13}$ OzGrav: The ARC Centre of Excellence for Gravitational Wave Discovery, Clayton VIC 3800, Australia \\
$^{14}$ Finnish Centre for Astronomy with ESO (FINCA), FI-20014 University of Turku, Finland \\
$^{15}$ Department of Astrophysics/IMAPP, Radboud University Nijmegen, P.O. Box 9010, 6500 GL Nijmegen, The Netherlands \\
$^{16}$ Departamento de Astrof\'{i}sica, Universidad de La Laguna, E-38206 La Laguna, Tenerife, Spain
}

\date{Accepted October 11 2021. Received October 8; in original form July 26}

\pubyear{2021}

\graphicspath{ {./images/} }

\begin{document}
\defcitealias{LVCGCNS190425z}{LVC 2019a}
\defcitealias{ligo2019ligo}{LVC 2019b}
\defcitealias{LVCGCNS190814bv}{LVC 2019c}
\newcommand{\Msun}{$M_{\odot}$}
\newcommand{\Lsun}{$L_{\odot}$}
\newcommand{\Rsun}{$R_{\odot}$}
\newcommand{\solar}{${\odot}$}
\newcommand{\mpcyr}{$\mathrm{Mpc}^{-3}\mathrm{yr}^{-1}$}
\newcommand{\lkyr}{$L_{\odot,K}^{-1} \mathrm{yr}^{-1}$}

\label{firstpage}
\pagerange{\pageref{firstpage}--\pageref{lastpage}}
\maketitle

\begin{abstract}
The Gravitational-wave Optical Transient Observer (GOTO) is an array of wide-field optical telescopes, designed to exploit new discoveries from the next generation of gravitational wave detectors (LIGO, Virgo, KAGRA), study rapidly evolving transients, and exploit multimessenger opportunities arising from neutrino and very high energy gamma-ray triggers. In addition to a rapid response mode, the array will also perform a sensitive, all-sky transient survey with few day cadence. 
The facility features a novel, modular design with multiple 40-cm wide-field reflectors on a single mount. 
In June 2017 the GOTO collaboration deployed the initial project prototype, with 4 telescope units, at the Roque de los Muchachos Observatory (ORM), La Palma, Canary Islands. 
Here we describe the deployment, commissioning, and performance of the prototype hardware, and discuss the impact of these findings on the final GOTO design. We also offer an initial assessment of the science prospects for the full GOTO facility that employs 32 telescope units across two sites. 
\end{abstract}




\section{Introduction}

The introduction of affordable large-scale CCDs, coupled with wide-field survey telescopes has transformed the detection rate of transients such as supernovae, extra-galactic novae, Galactic variable stars, outbursts from accreting binaries, and also near-earth asteroids. 
Amongst the most productive of these surveys are the the All-sky Automated Survey for Supernovae \citep[ASAS-SN;][]{shappee2014}, the Asteroid Terrestrial-impact Last Alert System \citep[ATLAS;][]{tonry18}, the Catalina Real-time Transient Survey \citep[CRTS;][]{Drake2009}, the Dark Energy Camera \citep[DECam;][]{Flaugher2015}, the Evryscope \citep[][]{Law2015}, HyperSuprimeCam \citep[HSC;][]{Aihara2018}, Pan-STARRS1 \cite[]{chambers16}, SkyMapper \citep[][]{Keller2007}, the Zwicky Transient Facility \citep[ZTF;][]{bellm19} and the upcoming BlackGEM array \citep[][]{Blackgem2015}. We also anticipate the addition of the Legacy Survey of Space and Time (LSST) at the Vera C. Rubin Observatory within the next few years \citep[][]{ivezi19}. 

The recent developments in wide-field all-sky optical surveys has been at least partly motivated by the increasing sensitivity of the Laser Interferometric Gravitational-wave Observatory (LIGO) and Virgo detectors \cite[]{advligo15, virgo15}. Due to their design, interferometric gravitational-wave (GW) instruments typically offer poor localisation accuracy, compared to traditional (electromagnetic) astronomical instruments. For a reconstructed GW signal, the sky localisation error region encompassing all possible signal origins can span many hundreds of square degrees \cite[e.g.][]{LIGORates}. The uncertainty arises primarily from the precision with which the signal arrival time delay can be measured, coupled with the relative signal strengths due to the different instrumental sensitivity patterns projected on the sky \citep[][]{Fairhurst2009}.

In order to maximise the chance of identifying an electromagnetic counterpart to a GW signal, follow-up instruments must promptly cover the maximum visible fraction of this sky region or, more accurately, the time-volume. 
This task is difficult for conventional optical telescopes, as their fields of view are usually measured in square arc minutes, requiring many individual pointings to cover the GW source localisation region. The use of alternative strategies, such as targeting individual galaxies within the region, which themselves can number in the hundreds to thousands also brings additional challenges \citep[e.g.][]{Ducoin2020, Gehrels2016}. For example, the GLADE catalog \citep[][]{GLADE2018} is complete only up to $D_L\sim 37$\,Mpc and uses luminosity as a tracer for the mass and merger rate of BNS sources. Consequentially, this strategy could result in missed events for those with low offsets from the host galaxy or those that originate in low-mass galaxies.

The Gravitational-wave Optical Transient Observer (GOTO\footnote{\url{https://goto-observatory.org}}) is an array of wide-field optical telescopes designed to efficiently survey the variable optical sky. It is specifically optimised for wide-field searches for electromagnetic counterparts to GW sources, complementing other search facilities and focusing on rapid identification of candidates.
Although not necessarily a typical event, the first binary neutron-star (BNS) merger, GW170817 validated many of the key design parameters of GOTO.
GW170817 was localised to within $\sim$28 square degrees of sky using LIGO and Virgo data \cite[]{gw170817, LIGORates}. The $V=16$~mag optical counterpart was discovered within $\sim$11~hr of the GW event followed by a lengthy multiwavelength campaign \citep[e.g.][]{Abbott2017,Australia2017,Arcavi2017,Chornock2017,sss17a,Covino2017,cowperthwaite17,Drout2017,Evans2017,Kasliwal2017,Lipunov2017,Nicholl2017,pian17,shappee2017,Troja2017,Utsumi2017,Valenti2017}, and its host galaxy was NGC~4993 at a distance of $\sim$40~Mpc \citep{LVC2017GCN,levan17,hjorth17}. 
Subsequent observations led to an avalanche of extraordinary observational data on an entirely new class of astrophysical event, providing insight into the production of short gamma-ray bursts \citep{gwgrb17,grb170817a,Savchenko2017,lyman18}, the origin of heavy elements \citep{pian17,smartt17,tanvir17} and even a new route to measuring cosmological expansion \citep{lvc17stdsiren,cantiello18}.
However, this event represents only the beginning of a new era of multi-messenger astronomy, and great diversity is to be expected as GW rates increase. Much is still uncertain around the physics driving the EM emission of mergers involving neutron stars. The EM luminosities, distances and source localisation properties will vary strongly between events and across science runs. Many of the key questions are still to be answered and this requires systematic efforts to identify and characterise these events. Early localisation is key, such that follow-up can unfold promptly. This need is the driving force behind the GOTO project.

In this paper we describe the design, deployment, commissioning, and performance of the GOTO prototype and look ahead towards the full deployment of the GOTO concept across two observing sites.
In \S\ref{sec:principles} we describe the principles informing the hardware design and specifications of the GOTO telescope system.
In \S\ref{sec:implementation} we describe the implementation, including the telescope control system, image processing pipelines, and observation scheduler, and assess their performance.
In \S\ref{sec:results} we describe the opportunities arising from survey and follow-up observations during the prototype commissioning, along with quantitative assessments of the instrument performance. Finally, in \S\ref{sec:conclusions} we assess the future prospects for detections of transients including the observational products of counterparts to binary neutron star inspirals.

\section{GOTO principles}
\label{sec:principles}

The GOTO concept was developed well before the first GW detections \cite[]{DarrenThesis}. The focus was a dedicated rapid-response system, targeting the early localisation of GW sources. 

At the time this goal presented significant challenges; not only were the early source locations expected to be very poorly constrained at the time of GW detection, there was also significant theoretical uncertainty for the electromagnetic properties of such events, including their luminosities as a function of energy and their decay timescales, among others. There are different strategies that one can take, reflecting a different balance between sensitivity, sky coverage and cadence. Our key design principles were flexibility, scalability and cost-effectiveness, with the GOTO instrumental capabilities tuned to complement other facilities suited for deeper observations and spectroscopic coverage.

We explored this parameter space of depth, area and cadence to find an optimal configuration. The GOTO hardware design centres on using arrays of relatively modest aperture, wide-field optical telescopes, hereafter referred to as unit telescopes (UTs), in order to survey the sky regularly in anticipation of detections. This approach was inspired by the SuperWASP approach to planet transit searching \cite[]{pollacco06}, which in turn inspired projects such as ASAS-SN.

There are two important factors for assessing the performance in this context, which define two distinct observing modes for the GOTO telescope system; ``triggered" and ``sky-survey" modes. First, the instrument must be able to respond promptly to a GW detection, targeting the specific areas on the sky that are consistent with the localisation constraints as provided by the multi-detector GW network.  In this response mode, hundreds to possibly thousands of square degrees need to be targeted, ideally with multiple visits, and fast enough to catch a short-lived source. 
Second, the instrument must be able to provide recent reference images (prior to the GW detection) with which to compare -- these would be acquired in a continuous all-sky survey mode. Although the difference imaging technique is a well-established tool in the variable star and transient community \citep[e.g.][]{alard98_diffimg,Alard2000} to remove the static foreground of sources effectively, many other variable and transient sources unrelated to the GW detection can be expected at any given time. The longer the time gap between triggered follow-up observations and the most recent sky-survey epoch(s), the more interlopers can enter, and it becomes increasingly more difficult to find the bona-fide object of interest. For this reason, one would want regular sky survey epochs, so that sources known to be variable prior to the GW detection can be discarded. Of particular relevance are supernovae, which are luminous for weeks to months. Over such large search areas significant numbers are visible at any given time. The combination of these two modes (both ``triggered" and ``sky-survey" modes) means that a large field of view is desired; the larger the field of view, the faster both modes are able to be completed.

As previously mentioned, the array approach offers a number of advantages. It allows the project to be scalable, with its capability set by the number of unit telescopes that can be deployed. An array also offers flexibility, as it can be deployed to maximise instantaneous field of view, depth at a more focused position, or provide different filters in individual telescopes. It is cost-effective, as the cost is linearly coupled to capability, and the implementation allows a good number of unit telescopes to be deployed at a site. 

A key constraint in this is the availability of cost-effective detectors. High-end professional large format CCDs would completely dominate the costs when employing large numbers of UTs, and would have complex cooling and electronics requirements. Our focus was instead on the much more affordable range of Kodak sensors, which offer exceptional price per pixel, albeit with a  reduction in the quantum efficiency (QE) as compared to high-grade devices. However, the cost reduction is so significant (order of magnitude), that it is then possible to consider using a significant number of cameras in order to make up for the loss of efficiency of a single camera. These types of sensors also perform well at relatively warm temperatures and therefore do not require sophisticated cooling systems.

With the pixel's physical size dictated by the sensor market, we then evaluated the performance of modest aperture telescopes using such sensors. Bigger apertures obviously improve the sensitivity. To make the most of the sensitivity, the optical design would need to sensibly sample the sensor pixels, with smaller pixel scales reducing the impact of sky background, but also reducing the achieved field of view. It is also desirable to be able to cycle through different filters such that both searching and characterisation can be optimized. The final constraint was the ability to multiplex without requiring a separate mount for each telescope. This is to reduce the physical footprint, complexity and cost of the facility. We pursued custom heavy-duty robotic mount systems capable of holding 4--8 telescopes at a time. 

We simulated a number of possible compromises, covering very wide-field configurations with 20~cm aperture telescopes, to more depth-focused options using fewer, larger telescopes. It was found that $D=40$~cm aperture unit telescopes were close to optimal, as this still allows us to multiplex the telescopes on a shared mount while offering a better depth/pixel scale compromise than smaller telescopes. A fast optical design would be needed to maximise the field of view, but also allow for a filterwheel. Multiple arrays of 8 telescopes could then cover the entire visible sky to moderate depths every few days, while multiple sites would ensure full sky coverage in both hemispheres. We present the implementation in more detail in the next section. We denote the ``prototype" as the 4 UT system (GOTO-4), the full-scale single-site system as GOTO-16, and the finalised full-scale dual-site observatory as GOTO-32.

\section{Implementation}
\label{sec:implementation}

\subsection{Hardware}
\label{sec:hardware}

As motivated in the previous section, the design of the GOTO telescopes was first and foremost driven by the sensor. In particular, the KAF-50100 CCD sensor produced by ON Semiconductor offered a very affordable large-format sensor, including 8304$\times$6220 pixels at a scale of 6~$\mu$m. The sensor was also offered in a convenient compact package by Finger Lakes Instrumentation (FLI) as part of their MicroLine range (ML50100). We provide more details on the detector performance in \S \ref{sec:detectors}.

\begin{figure}
	\includegraphics[width=\columnwidth]{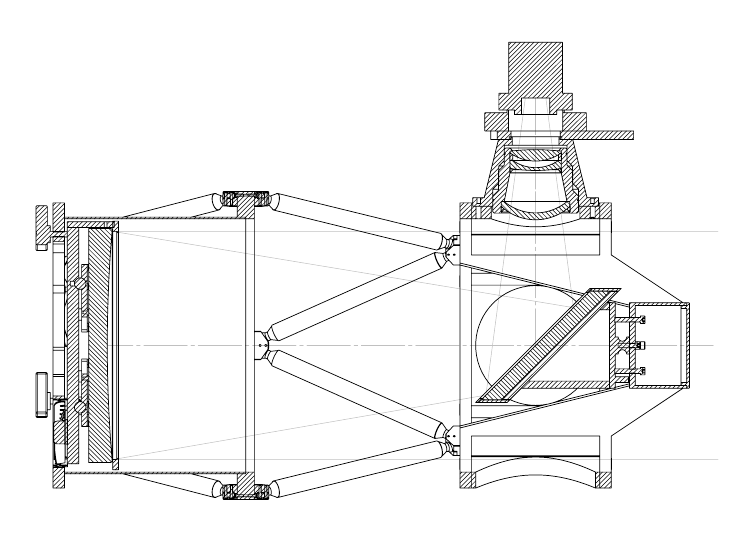}
    \caption{The GOTO prototype unit telescopes make use of a Wynne corrector in a Newtonian configuration. The $D=40$\,cm f/3 primary mirror has a hyperbolic surface and is supported by a mirror cell that allows for three-point collimation adjustment. An elliptical secondary direct the lights towards the multi-lens corrector system that projects a collimated effective f/2.5 beam. The instrumentation is mounted off-axis with a stage of tip-tilt, a robotic focuser, a 5 slot filter wheel and the camera enclosure. The structural support is provided by a carbon-fibre open truss arrangement. 
    \label{fig:APMtube} }
\end{figure}

\begin{figure}
	\includegraphics[width=\columnwidth]{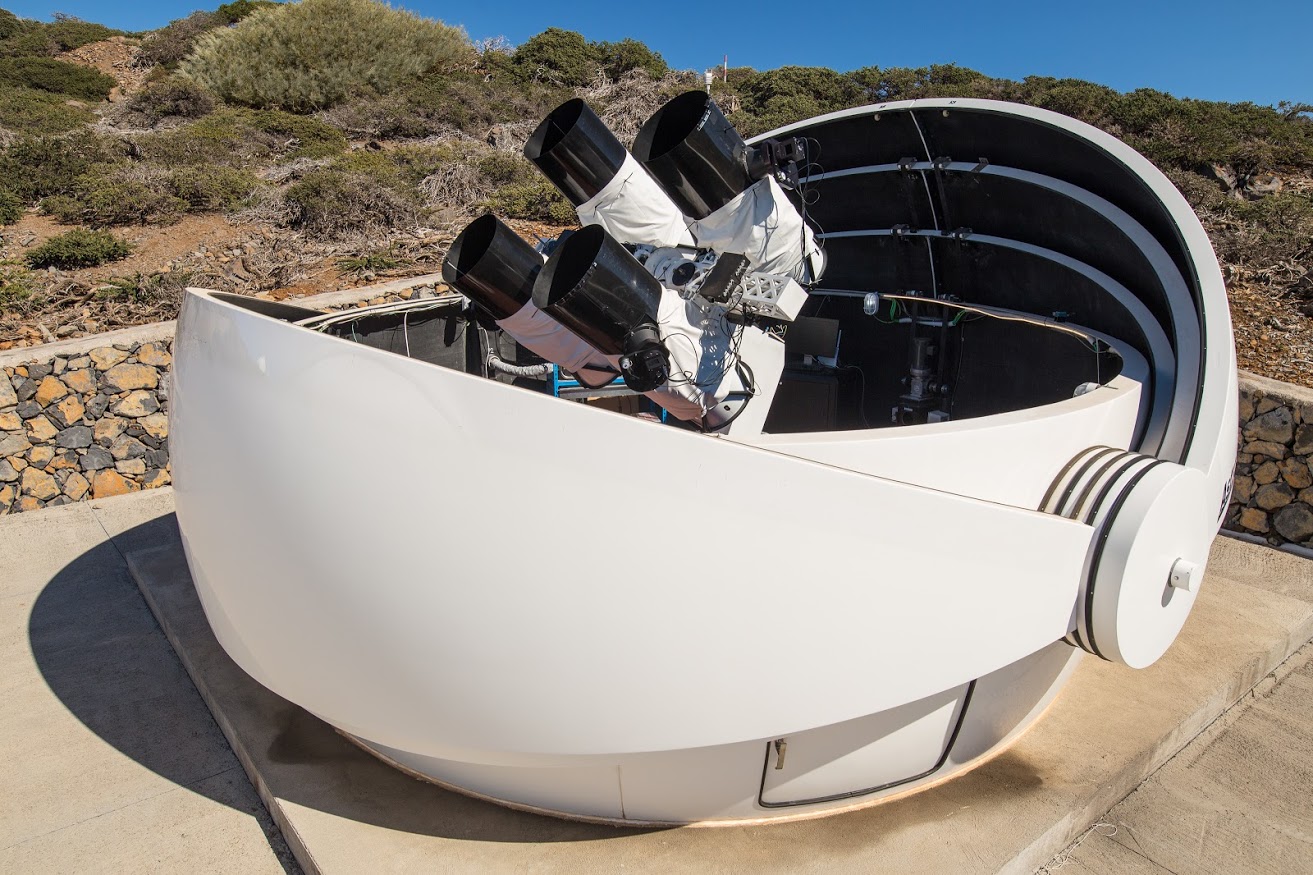}
    \caption{A photo of the GOTO-4 prototype telescope system at the Roque de los Muchachos Observatory in 2018, loaded with the initial 4 prototype unit telescopes inside of an 18-ft Astrohaven dome.}
    \label{fig:photo}
\end{figure}

\begin{figure}
	\includegraphics[width=\columnwidth]{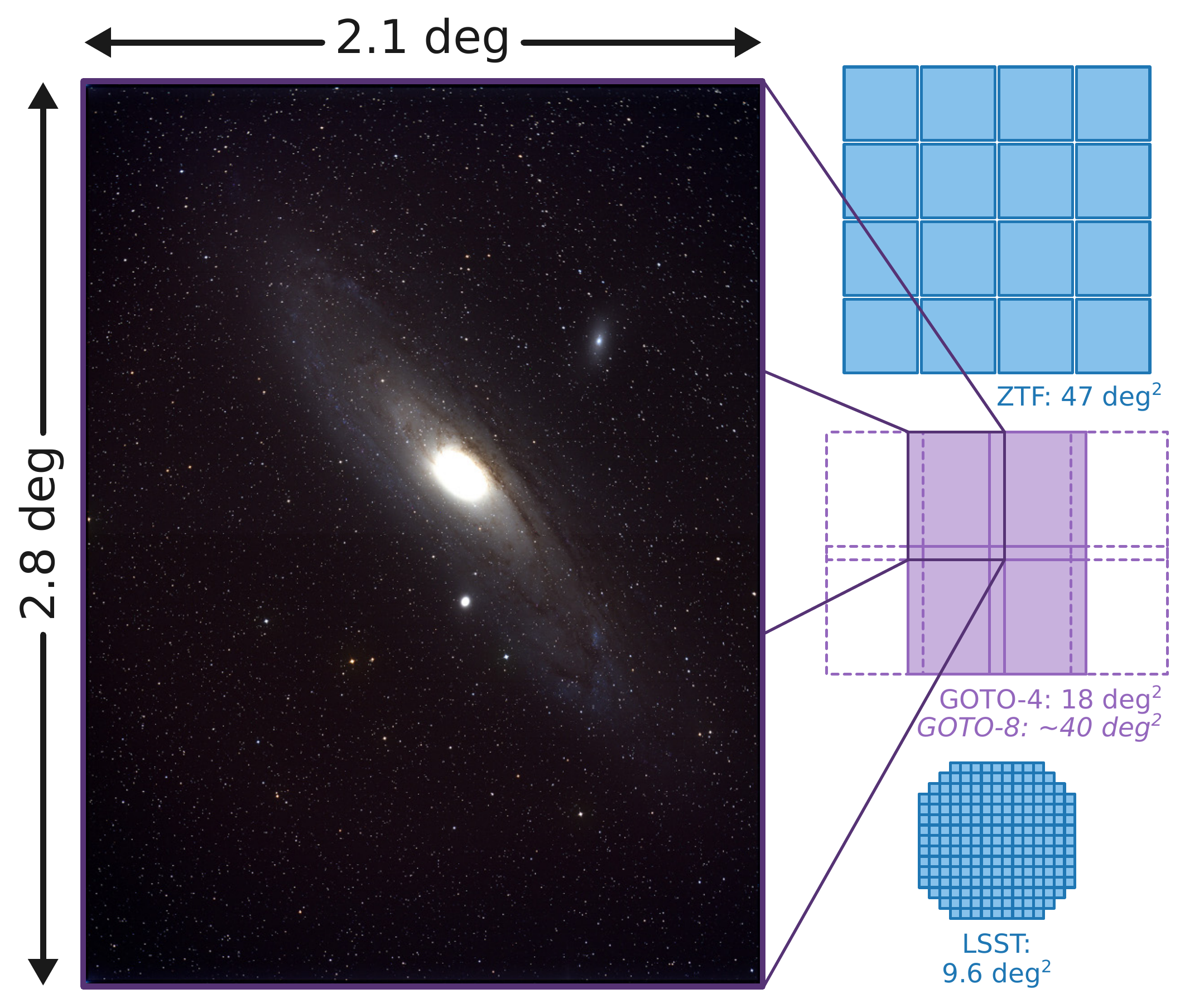}
    \caption{The GOTO prototype field of view. On the left is a commissioning image of M31 taken with one of GOTO's cameras, showing the wide field of view of a single unit telescope. Four unit telescopes together create the initial 18~square~degree prototype survey tile (GOTO-4), which will increase to 40~square~degrees in the full GOTO-8 system (shown by the dashed boxes). For comparison, the fields of view of two other wide-field projects are shown to scale on the right: the Zwicky Transient Facility and the Rubin Observatory LSST Camera.}
    \label{fig:fov} 
\end{figure}

In order to provide a sensible pixel-scale, the prototype optical tube assemblies (OTAs) for the GOTO UTs were designed to offer an aperture of $D=40$~cm at f/2.5 (Fig.~\ref{fig:APMtube}). This maps to 1.25 arcsec per pixel, small enough to control sky background yet critically sampling the point-spread function (PSF) and offering a field of view of $\sim$5 square degrees. To deliver a corrected field, the design deploys a set of corrector lenses in between the secondary mirror and the focal plane. As it was desirable to be able to deploy filters, the optical design is Newtonian, allowing for a traditional filter wheel at the Newtonian focus. In our case, we coupled a 5-slot FLI filter wheel (CFW9-5) to the FLI camera package. The initial set of filters were the Baader set, which offers three colour bands ($R$,$G$,$B$) as well as a wide-band $L$ filter (see section~\ref{sec:sensitivity}).

The first phase of the GOTO project involved the development and construction of a prototype telescope, with 4 UTs mounted on a custom robotic mount (see Fig.~\ref{fig:photo}). The mount is a German equatorial design, and the unit telescopes are loaded symmetrically to keep the system balanced.
The mount drive used a wormwheel implementation where the two axes motors transfer torque to the mount wheels via a worm-gear. The gear is tensioned to push into the wormwheels but can decouple under overload for safety. The tension can be adjusted to find a balance between stiffness of the gear versus the ability to slew smoothly under load without overloading the motors.
Encoders on the motors and high-resolution Renishaw encoders on the two axes permit accurate active dual encoder mount position control.
Steel boom arms protrude to both the East and the West side to accommodate the tubes, control electronics, control computers and balance weights.
Each unit telescope is connected to the mount boom arm via an adjustable guidemount, which allows individual UTs to be rotated and tilted ($\pm$5 deg) so that the footprint of the combined array can be defined. In the prototype configuration the entire array covers 18.1~square~degrees in a single pointing (see Fig.~\ref{fig:fov}). The field of view of individual unit telescopes intentionally overlap, to provide a contiguous field of view which allows for effective tiling on the sky without gaps. The overlap regions also provides important cross-calibration checks for the pipeline. In principle the guidemount adjustment range is sufficient to allow all the unit telescopes to co-align, or be arranged into more complex shapes, but the default arrangement allows a wider combined field of view and therefore prioritises sky coverage.

Whilst the prototype phase only included 4 UTs (2 on either side), the mount was designed from the start to be able to hold 8 UTs. A complete mount array would produce a field of view of $\sim35-40$~square~degrees, as shown in Fig.~\ref{fig:fov}, comparable to the 47~square~degree field of view of ZTF. 
In order to deliver full sky coverage and a cadence of a few days, it was envisaged that four of these full 8 UT arrays can then be located across the globe at two sites in opposite hemispheres to achieve the targets outlined in Section~\ref{sec:principles}. Spreading four 8-UT arrays over two sites (rather than four locations) was done to alleviate logistical and infrastructure challenges that come with setting up and operating at each location. This setup would result in an instantaneous field of view of up to $\sim$80~square~degrees at each site, split across two mounts and, given proper choice of sites, provide near 24-hour coverage for a fraction of the sky and coverage of all declinations.

The prototype telescope was deployed at the Roque de los Muchachos Observatory, La Palma, the intended home of the first GOTO site and a premier observing site in the Northern hemisphere. The GOTO site is operated by the University of Warwick on behalf of the GOTO consortium and was funded by the founding members. The system is housed in an Astrohaven 18ft clamshell dome enclosure, offering panoramic access to the local sky down to 30 degrees altitude. Additional customisations were added to the dome to facilitate secure robotic operations, including extra sensors, in-dome cameras and sirens \citep{dyer18}.

The key goal of the GOTO prototype was to demonstrate the viability of the design choices before scaling the project up with additional telescopes. We also wanted to deploy it timely enough to ensure that the prototype could pursue actual GW searches during the advanced LIGO-Virgo observing runs. The prototype achieved first light in June 2017, followed by its official inauguration in July 2017. A summary of the key specifications are provided in Table~\ref{tab:specs}.

\begin{table}
	\centering
	\caption{GOTO prototype hardware specifications.}
	\label{tab:specs}
	\begin{tabular}{lr}
		\hline
        \textbf{Parameter} & \textbf{Value}\\
        \hline
        \multicolumn{2}{c}{\textbf{Site}} \\
		Latitude & $28\degr45'36.2''$ N \\
		Longitude & $17\degr52'45.4''$ W \\
		Altitude  & 2300~m a.s.l. \\
		Dome design & Clamshell \\
        Dome diameter & 18~ft (5.5~m) \\
        
        \multicolumn{2}{c}{\textbf{Mount}} \\
        Mount design & German equatorial (parallactic) \\
		Mount slew rate & 4--5~deg\,s$^{-1}$ \\
        UTs per mount & 8 (4 filled) \\
        
        \multicolumn{2}{c}{\textbf{Unit telescopes}} \\
        OTA design & Wynne-Riccardi \\
        Primary diameter & 40~cm \\
        Primary conic constant & -1.5 \\
        Secondary diameter & 19~cm (short axis)\\
        Secondary conic constant & N/A (flat) \\
        Corrector diameter & 12~cm \\
        Focal ratio & f/2.5 \\
		Field of View & 2.1~deg $\times$ 2.8~deg\\
        \multicolumn{2}{c}{\textbf{Detectors}} \\
        Detector size & 8304 $\times$ 6220 pixels \\
        Active region & 8176 $\times$ 6132 pixels \\
        Pixel size & 6~$\mu$m \\
		Pixel scale & $1.25''$/pixel\\
		Filters & Baader $R$, $G$, $B$, $L$ \\
		Gain & 0.53 -- 0.63 $e^{-}$/ADU \\
		Readout Noise & 12 $e^{-}$ \\
		Dark current noise & < 0.002 $e^{-}$/s \\
		Full-well capacity & 40300 $e^{-}$\\
		Fixed-pattern noise & 0.4\% full-well capacity \\
		Non-linearity & < 0.2\% \\
		\hline
	\end{tabular}
\end{table}

\subsection{Software}
\label{sec:software}

The GOTO software was developed in-house and is divided into multiple components, each of which is described in the sections below. Almost all of the GOTO software was written in Python and makes use of Python-based packages.

\subsubsection{Robotic telescope control}
\label{sec:gtecs}

GOTO operates using a custom control system, G\nobreakdash-TeCS (the GOTO Telescope Control System; \citealt{dyer18}, \citealt{dyer20}, \citealt{DyerThesis}). G\nobreakdash-TeCS is written in Python and is based on the code developed for \textit{pt5m} \cite[]{pt5m}.

\begin{figure*}
	\includegraphics[width=\linewidth]{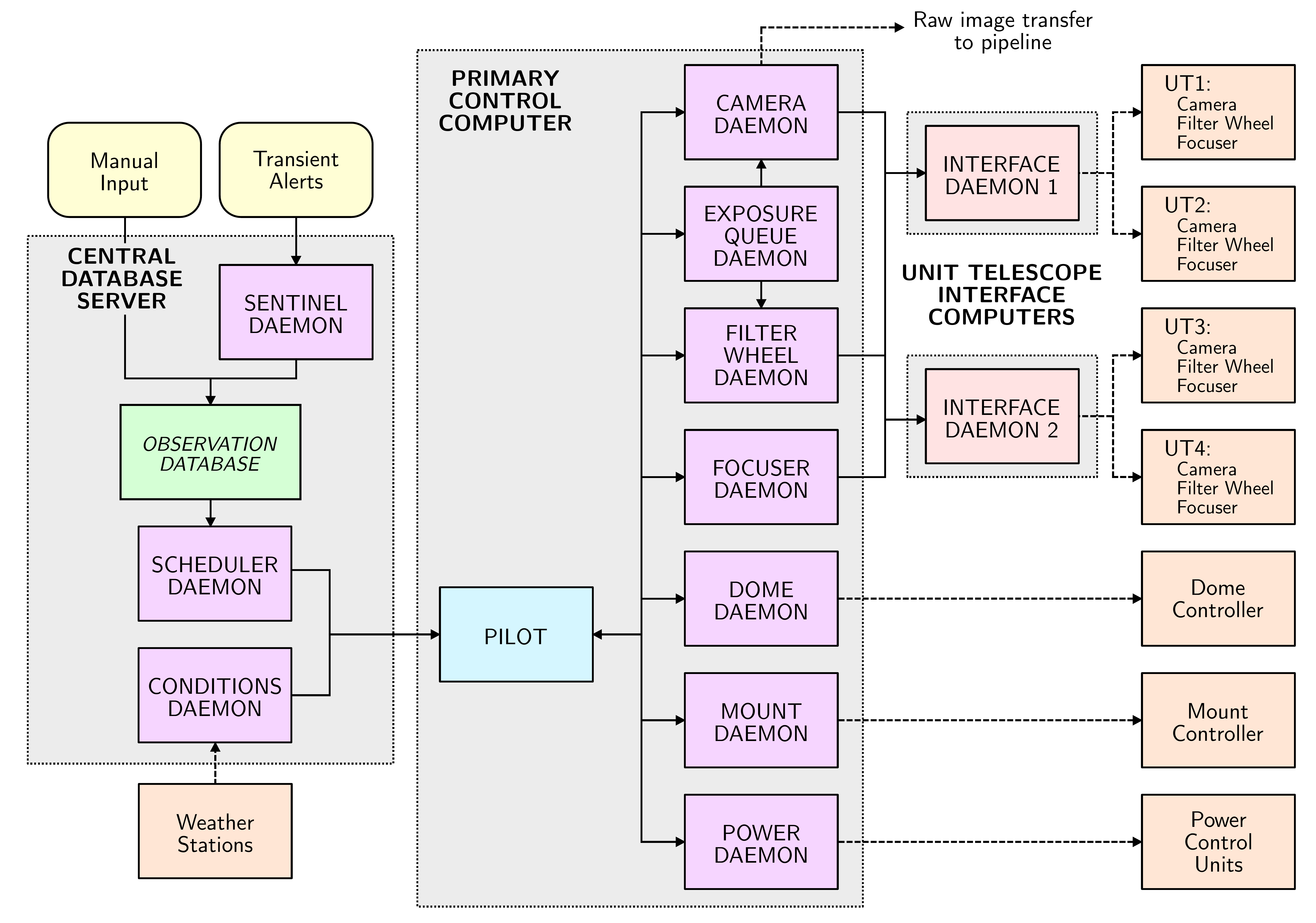}
    \caption{The G-TeCS software architecture. The observation database along with the sentinel, scheduler and conditions daemons are located on a central server (left). The pilot and hardware daemons for the telescope are run on the primary control computer located in the GOTO dome (centre). The hardware daemons communicate with their respective hardware units (right) directly or via interface daemons (in the case of the unit telescopes).}
    \label{fig:gtecs}
\end{figure*}

The primary software programs within G\nobreakdash-TeCS are a series of daemons; background processes that monitor and provide an interface to their hardware units. The daemons interact using the Python Remote Objects (Pyro) module\footnote{\url{https://pythonhosted.org/Pyro4/}}; each daemon is a Pyro server which allows communications between processes and daemons across the local network. Figure~\ref{fig:gtecs} shows a schematic view of the G\nobreakdash-TeCS software architecture.

There are six primary hardware daemons each named after the category of hardware they control: the \emph{camera}, \emph{filter wheel}, \emph{focuser}, \emph{dome}, \emph{mount} and \emph{power} daemons. These are run on the primary control computer located within a rack in the GOTO dome. Due to GOTO's array design the unit telescope hardware (the cameras, focusers and filter wheels attached to each UT) are connected in pairs to interface computers mounted on the boom arm. Each category of hardware is then controlled in parallel by their respective daemons running on the primary control computer. A seventh hardware daemon, the \emph{exposure queue} daemon, processes sets of exposures and handles timing between the camera and filter wheel daemons, allowing sets of exposures to be observed in sequence and ensuring that the correct filters are set before each begins.

Three additional support daemons run on a central server alongside the primary observation database, located on La Palma in the neighbouring SuperWASP telescope enclosure. The \emph{sentinel} daemon processes incoming transient alerts and adds targets to the database, which are then processed and sorted by the \emph{scheduler} daemon to determine the highest priority target to observe at the given time (see \S~\ref{sec:scheduling}). In addition the \emph{conditions} daemon collects and processes data from the on-site weather stations in order to determine if it is currently safe to open the dome.

To enable GOTO to function as a fully robotic telescope the daemons are issued commands by the \emph{pilot} control program, which acts in place of an on-site human operator. The pilot is an asynchronous Python script that runs through a series of tasks every night: powering up the system in the late afternoon, taking bias and dark images, opening the dome after sunset, taking flat fields and focusing the telescopes, observing targets provided by the scheduler daemon throughout the night, taking flat fields again in the morning twilight, and finally closing the dome and shutting down the system at sunrise. Throughout the night the pilot monitors the local weather conditions reported by the conditions daemon, as well as the status of the telescope hardware. If the conditions are reported as bad then the dome will close and the pilot will pause until they are clear. If a problem with the hardware is detected then the pilot will run through a series of pre-defined recovery commands in order to try and repair the system; if these fix the problem then the pilot will resume observations, but if the error persists then the pilot will issue an alert before shutting down. During the night the pilot sends messages to a dedicated channel on Slack\footnote{\url{https://slack.com}}, a messaging application workspace, both regularly scheduled reports (a weather report in the evening, a list of observed targets in the morning) as well as alerts for any errors that might require human intervention. The control system can also be switched over to manual mode if desired, pausing the pilot and allowing a remote observer control of the telescope.

The G\nobreakdash-TeCS architecture has been designed to be modular and the overall system is easily expandable. For instance, adding the second set of four unit telescopes to the prototype only requires new interface daemons, which are then integrated into the existing system. In the future as more GOTO telescopes are commissioned each array will be controlled by an independent pilot, which will receive targets from a single central scheduler. This will allow a rapid, coordinated response to any transient alerts.

\subsubsection{Observation scheduling}
\label{sec:scheduling}

As a survey telescope, GOTO observes target fields aligned to a fixed all-sky grid, to ensure consistently-aligned frames for difference imaging. For the GOTO-4 prototype this grid is formed of tiles with a size of 3.7~degrees in the right ascension direction and 4.9~degrees in the declination direction, combining the field of view of all four cameras into a single 18.1~square~degree field with some overlap between the neighbouring cameras (as shown in Fig.~\ref{fig:fov}). The all-sky grid is defined by dividing the sky into a series of equally spaced 18.1~square~degree tiles; 2913 tiles in total cover the entire celestial sphere. Just over 700 tiles are visible at any one time when considering the local horizon, and approximately 76 per cent of the celestial sphere is visible over the course of a year from GOTO's site on La Palma (see Fig.~\ref{fig:tile_counts} in \S~\ref{sec:coverage}).

The G\nobreakdash-TeCS sentinel daemon contains the system alert listener, which monitors the NASA GCN \citep[Gamma-ray Coordination Network][]{GCN} stream for relevant astrophysical events. During the prototype phase GOTO-4 responded to gravitational-wave alerts from the LIGO-Virgo Collaboration (LVC; see \S~\ref{sec:gw}), as well as gamma-ray burst (GRB) events from the \textit{Fermi} Gamma-ray Burst Monitor \citep[GBM;][]{Meegan2009} and \textit{Swift} Burst Alert Telescope \citep[BAT;][see \S~\ref{sec:grbs}]{Krimm2013}. When one of these alerts is received by the sentinel, the skymap containing the localisation region is mapped onto the predefined all-sky grid, in order to find the contained probability within each tile. These tiles are then inserted into the observation database in order of probability until the entire 90 per cent localisation region has been covered.

In order to determine which of the targets in the database to observe, the scheduler daemon first applies several observing constraints on the queue of pending pointings using the \texttt{astroplan} Python module \citep{astroplan18}. The constraints include checking the target's altitude above the local artificial horizon, the distance of the target from the Moon and the current lunar phase (targets can be limited to bright, grey or dark time). Once invalid pointings have been filtered from the queue, those remaining are sorted by the rank defined when they were inserted into the database. Gravitational-wave follow-up pointings rank higher than those from GRB alerts, and both are always higher than normal survey pointings. For events with large skymaps spanning multiple tiles (such as almost all gravitational-wave detections so far) the pointings that are yet to be observed are prioritised over repeat visits of previously-observed tiles, ensuring that the visible localisation region is covered rapidly. For any pointings that are still equally ranked a tiebreak parameter is constructed based on the skymap localisation probability contained within the tile and the current airmass of the target, to prioritise both covering the high-probability regions of the skymap and data quality. The resulting target with the highest priority is returned to the pilot to observe. This ranking system functions as a ``just-in-time'' scheduler; the pilot queries the scheduler every 10 seconds, the scheduler then recalculates the pointings queue and returns the pointing that is currently the highest priority. This results in a system that is very quick to react to transient alerts, as new targets added to the database are automatically sorted at the top of the queue, and GOTO has been able to begin observations of new events within 30 seconds of the alert being received by the sentinel (see \S~\ref{sec:results}).

\subsubsection{Image processing}
\label{sec:pipeline}

No significant image processing is performed on La Palma. For each observation, images from each camera are saved as individual frames by the G\nobreakdash-TeCS camera daemon using the FITS (Flexible Image Transport System) format and are then compressed and transferred to a data centre based on the campus of Warwick University (Coventry, UK). A dedicated level-2 VLAN fibre connection was set up for this purpose, providing a secure 1\,Gb connection between the observatory and the campus. This connection provides ample bandwidth to transfer images while the next set is being exposed, and should allow real-time processing even when the envisaged full-site of unit telescopes will be exposing in parallel.

\begin{figure}
	\includegraphics[width=\columnwidth]{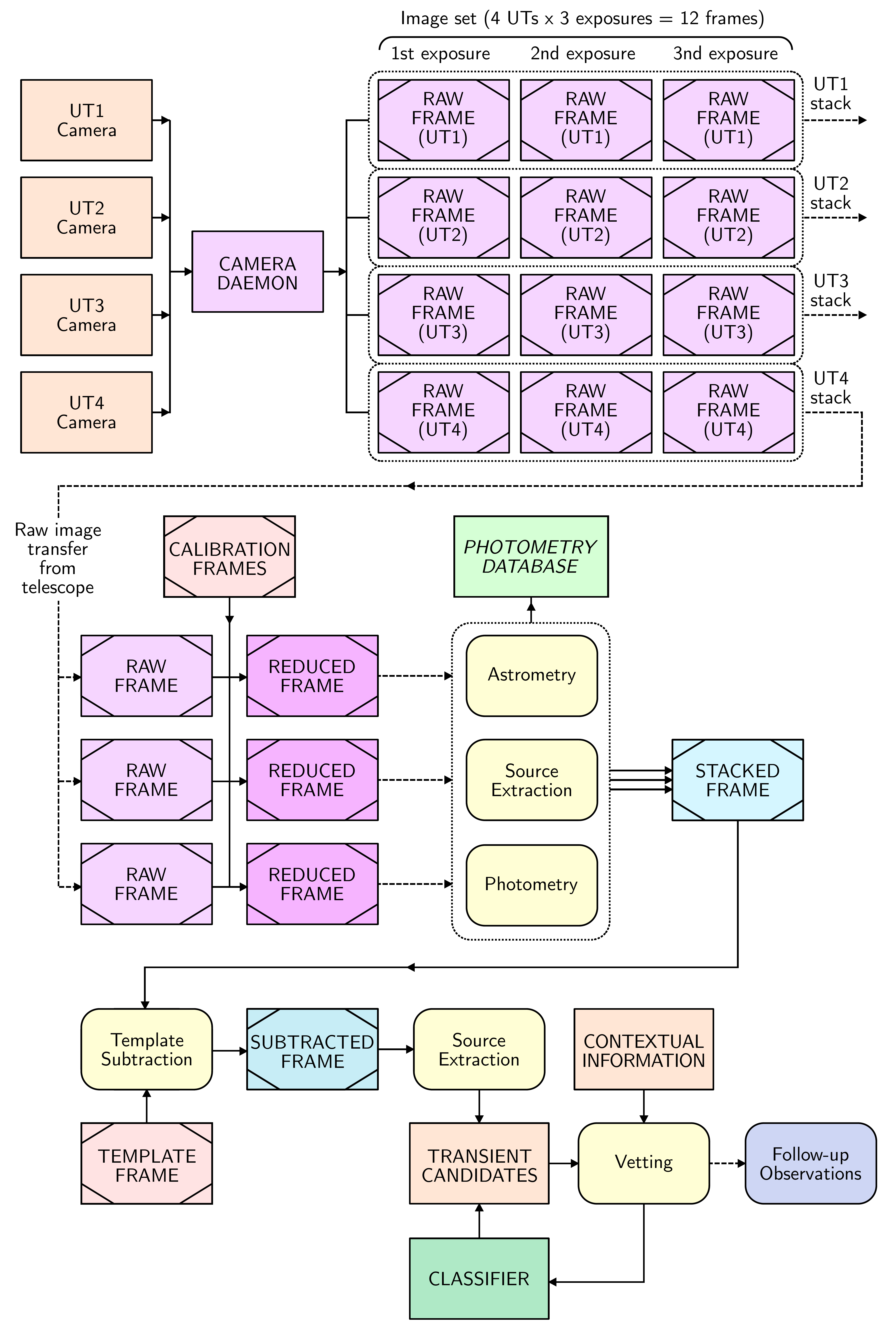}
    \caption{Data processing flow. For a set of three exposures 12 individual frames are produced: a stack of three from each of the four UTs. Each stack is processed in parallel, first with each raw frame being calibrated to produce a reduced frame. These reduced frames are stacked as a group where astrometry, source extraction and photometry are performed. The entire photometric catalogs are stored in the photometry database and added to the FITS image. The median stacked frame is matched with a stacked template image from the template database and subtracted using {\sc hotpants}. Once subtracted, a list of transient candidates are sent out to be vetted. The vetting process in its final stage is manual where contextual information about the source is provided as well as the the classification score from the real-bogus classifier. If any candidates have passed all vetting stages, then they are sent to other follow-up facilities for further characterisation. }
    \label{fig:pipeline}
\end{figure}

A watcher script in the data centre monitors the arrival of new data files and adds them to the queue for processing with the prototype data reduction pipeline, {\sc GOTOphoto} (Fig. \ref{fig:pipeline}). The data centre hardware is a dedicated stack of high-performance server nodes, with some dedicated to offer NAS storage while others serving as database servers, and a group of identical compute nodes for processing. The stack is on a local 10\,Gb interconnect throughout and also links to other campus subnets at 10\,Gb.

The data-flow is designed to allow real-time data processing with low latency. The initial stages perform standard CCD bias, dark, and flat-field corrections for each science frame. The corrections are performed using calibration files from deep stacks of frames taken across multiple nights as a more robust and reliable method than using nightly stacks. A source detection pass using {\sc SExtractor} \citep{sextractor} is then made, identifying the locations and performing preliminary instrumental photometry for sources in the frame. An initial astrometric solution is then found using {\sc astrometry.net} \citep{lang10} with their pre-built {\em Gaia} indices. The fitting process uses the telescope pointing as a starting point to search in right ascension and declination, and fixes the pixel scale to that of each telescope. Although the fast optics suffer from significant distortions across the field of view, the large number of point sources available in each frame offer good constraints for the astrometry. The quality of this initial solution is then checked, and the higher order terms further refined if necessary. This refinement uses our principal reference catalogue, ATLAS-REFCAT2 \citep{atlasrefcat2}, for cross-matching. A custom package\footnote{\url{https://github.com/GOTO-OBS/goto-astromtools}} is used to iteratively refine the SIP (Simple Imaging Polynomial) distortion parameters of the WCS (World Coordinate System) solution for improving the sky to frame coordinate transformation, updating the linear and polynomial coefficients sequentially to ensure stable convergence. More robust quality flags are computed using the reference catalogue, applying information about the local quality of the astrometric solution to the source tables. The quality flag is a combination of bit values indicating whether parameters such as the astrometric solution or the mean full width at half maximum of the stellar profiles are significantly greater than the expected values. These flags take binary values up to 128, with the most severe defects attracting higher values. After refitting, the typical astrometric RMS noise in each frame is $\sim$0.6 arcsec (or less than half of the detector's pixel scale). The cross-matched reference catalogue is then used to calibrate the initial instrumental photometry found earlier. Kron apertures \citep[][]{Kron1980} are used for measurements of all sources in the frames with a typical baseline calibration uncertainty of 0.03 mag. 

After the above processing, an individual science frame is considered finished. Further stages of the data-flow rely on small stacks of these individual frames, which form exposure sets. The scheduler almost exclusively employs an observing strategy where multiple exposures are obtained at each pointing for increasing the S/N of each set -- a set is typically 3-4 exposures, each 30-90 seconds long. A processing queue is aware of the assignment of individual frames to a given set using header cards denoting the total number of frames to be included in the set, and the position of the current frame in that set. Once a set has had all its individual frames processed, they are aligned and median-combined. Given the typically small alignments required between frames, the alignment procedure is a simple translation of the frames, fixing rotation, scale and higher order terms. Combination is done via a relatively na\"ive scaled-median approach. After this, the stack is sent through the same source detection, astrometry and photometry routines as was done for the frames, to produce the final science image for the pointing. We note that various phenomena can cause an abrupt end to a set of exposures, e.g. weather or target-of-opportunity override. In these cases the pipeline has a default wait period, of order an hour, after which it considers a set finished, regardless of whether the expected number of exposures matches those that were processed, and the partial stack is sent forward for processing as above.

\subsubsection{Template images} 
Science frames undergo additional standard difference-imaging processing as a means to identify variable and new objects in the fields of view. A `template bank' of observations of tile pointings is maintained, which are generated from historical visits to a given tile and using the best quality frame available (determined from a combination of PSF characteristics and limiting magnitude of the frame). This template bank is searched for a suitable template frame from which to subtract a given set science frame and principally matching on the UT, filter and coordinates on sky. Given the distortions across the fields of view, and the consequential requirement for significant alignment, including arcminutes translations, rotation and the need for high-order transformation terms, we employ our own customised alignment routine {\sc spalipy}\footnote{\url{https://github.com/Lyalpha/spalipy}}. Briefly, the routine finds an initial affine transformation between two matching ``quads" \citep{lang10} of stars between sets of frames and fits a smooth 2D spline surface to the x- and y-pixel residuals between cross-matched sources. The 2D spline surface is applied to the final transformation and robustly handles non-homogeneous coordinate mapping to align the science and template frames to within a sub-pixel accuracy. The aligned template frame is subtracted from the science frame using {\sc hotpants} \citep{hotpants} to produce a difference frame. Finally, this difference frame is then passed through {\sc SExtractor} to identify sources (see \S \ref{sec:sources}).

\subsubsection{Database}
\label{sec:database}

The valuable metadata, including photometry, for a processed science frame is stored in the header and various FITS table extensions of its file. However, for ease and speed of access, this data is also stored in a Postgres database (DB) held on a dedicated server node. Since at the core of most queries is some reliance on sky coordinates (whether searching for images covering a particular location, or cross-matching photometry to create light curves), indexes are generated for \texttt{ra}- and \texttt{dec}-like columns using the {\sc q3c} Postgres extension \citep{q3c}.

The ATLAS-REFCAT2 is also stored as a {\sc q3c}-indexed Postgres table, which is queried as part of the data-flow (\S \ref{sec:pipeline}). Performance of the DB is heavily optimised by Postgres and makes use of the sizeable cache available from the 128 GB memory on the current DB server. As such, query speeds can be variable, but, as an example, returning sources in a typical GOTO UT field of view ($\sim10^{4}-10^{5}$ rows) from the total $\sim10^9$ sources in the ATLAS-REFCAT2 catalogue takes less than a few seconds, and substantially less than one second if a similar query has been performed recently (which is often the case when processing frames from exposure sets taken at the same sky position).

\subsubsection{Transient \& Variable Source identification}
\label{sec:sources}

In order to identify transient and variable sources, difference imaging is employed (\S \ref{sec:pipeline}). Such difference imaging does not provide a clean representation of the new or varying sources in the field alone. Since the image subtraction algorithm must handle varying levels of image-depth and PSF shapes, subtraction residuals are almost entirely unavoidable \citep{alard98_diffimg, Alard2000, zackay16_zogy, masci17_iptf_pipeline}. These residuals often appear as valid source detections to most algorithms (including {\sc SExtractor}, used here), and they generally far outnumber any astrophysically real detections in the difference frame. In order to elucidate the objects of interest, various methods involving machine learning have been pioneered to calculate probabilistic scores for the detections. These scores are often described on a scale of ``real" to ``bogus", giving rise to the ``realbogus'' name to describe such models. The models can then be used to filter out image-level contaminants, such as spurious residuals and related CCD artefacts, in the difference frames \citep{brink13, wright15, duev19}.

The early version of the GOTO data-flow employed a Random Forest (RF) model which matched quite closely the one presented in \citet{bloom12}. However, the significant optical distortions of the UTs meant that the difference images were particularly challenging for the model in most cases. The lack of historical GOTO data meant training the supervised model was also difficult and had to rely on fake source injections to produce sufficient ``real" sources. This meant properly characterising its performance was also difficult. To overcome this we generated a much improved model, using instead a convolutional neural network (CNN) to analyse the pixel-level data (in contrast to extracting human-selected ``features" of the detections, as is required for the RF approach), and harvested very large samples of ``real" and ``bogus" sources from actual data, with novel augmentation techniques to improve the recovery of various types of transients and across a whole variety of observing conditions. A preliminary version of this approach was implemented in July 2020 and resulted in drastically improved recovery of transients when compared to external streams (such as spectroscopically-confirmed Transient Name Server objects). For a fixed false positive rate of 1 per cent, the newly-implemented classifier achieved a 1.5 per cent false negative rate on a held-out test set, and reached a $\sim$97 per cent recovery rate when evaluated on a benchmark dataset of real observations of confirmed transients. The CNN model and the automated data-generation techniques are described fully in \cite{killestein21_gotorb}.

\begin{figure*}
	\includegraphics[width=\linewidth]{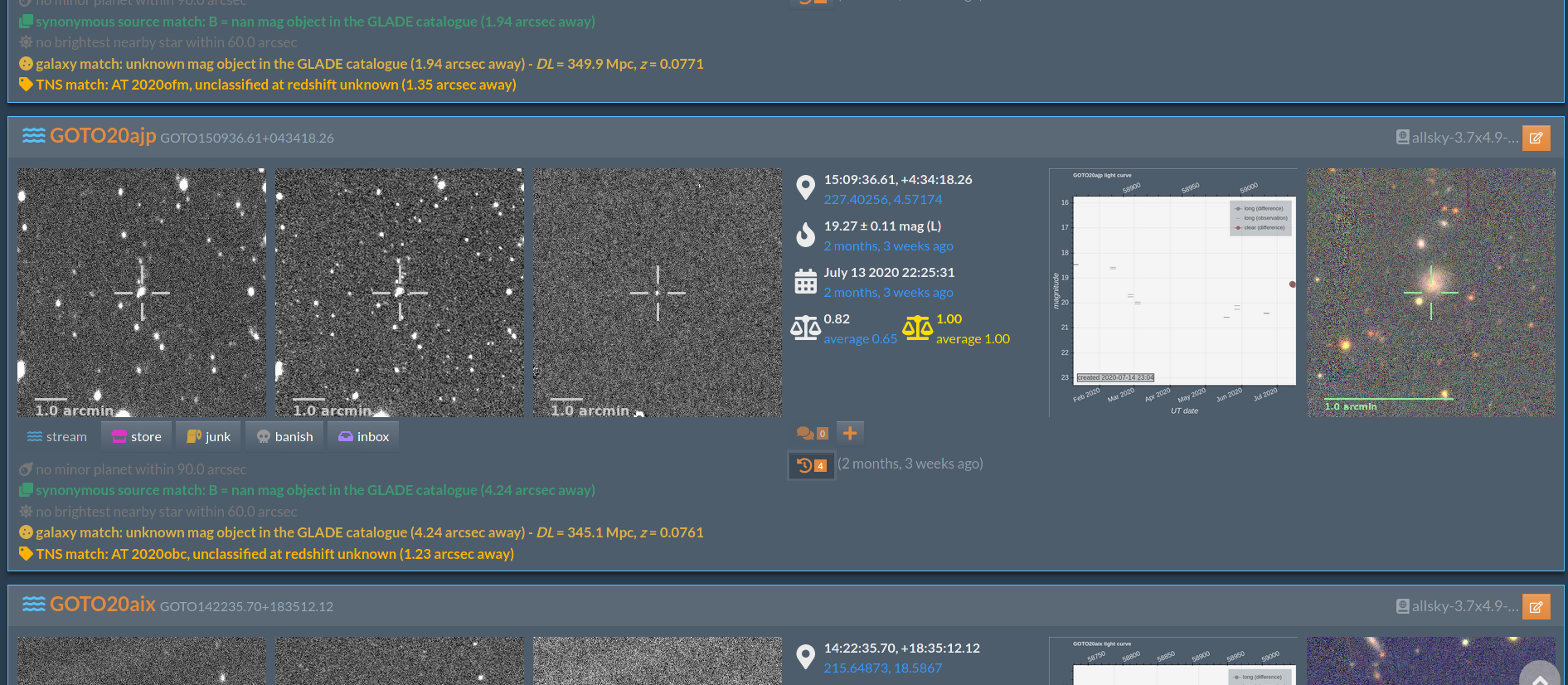}
    \caption{An example screenshot from the GOTO Marshall web interface. Shown is a list of source tickets providing at-a-glance information for each new source that passes preliminary cuts on the real-bogus score. Links within the ticket can take the user to pages showing more information on the source and its photometry. Users are also able to comment and provide additional classification for the sources, as well as assigns them to their own (or shared) ``watchlists". }
    \label{fig:marshall}
\end{figure*}

Once difference frame sources have been scored, they are presented to end-users via a web interface ``Marshall" (a screen shot of which is shown in Figure \ref{fig:marshall}). The GOTO Marshall is powered by the {\sc django}\footnote{\url{https://www.djangoproject.com/}} web-framework utilising its own Postgres DB backend, and exploiting {\sc celery}\footnote{\url{https://github.com/celery/celery}} to manage its internal tasks. At regular intervals a {\sc celery} task scrapes the \texttt{candidate} table of the {\sc GOTOphoto} for new rows that pass some threshold on the classifier's real-bogus score. The ingestion of a new entry generates a cascade of tasks to aid end-users in their decision on the scientific merit of a source, such as generating image-stamps and light curve plots, and perform contextual information checking on the source by cross-matching with astrophysical catalogues (through the {\sc catsHTM} interface, \citealt{catshtm}), and minor planet ephemerides.

\subsection{Performance}
\label{sec:performance}

\begin{table*}
\centering
\caption{Zeropoint calibration performance for the GOTO-4 prototype system under dark lunar conditions. The airmass-corrected calibration is completed against the APASS survey for each frame and the performance is calculated against the expected theoretical magnitudes. The zeropoint performance is measured as $10^{(ZP - ZP_\mathrm{model})/2.5}$. The expected 5$\sigma$ limiting magnitudes are given using $t$=60\,s observations under dark (D), grey (G), and bright (B) conditions.}
    \label{tab:performance}
    \resizebox{\linewidth}{!}{
\begin{tabular}{cccccccccc}
\hline \hline
Telescope & Filter & APASS filter & Model Extinction & Model ZP & Real ZP & Performance & 5$\sigma$ Lim Mag (D) & 5$\sigma$ Lim Mag (G) & 5$\sigma$ Lim Mag (B) \\ \hline
UT1 & $L$ & $g'$ (AB) & 0.108 & 22.63 & 22.47 & 86\% & 19.80 & 19.54 & 19.35 \\
UT1 & $R$ & $r'$ (AB) & 0.063 & 21.33 & 21.27 & 94\% & 18.59 & 18.50 & 18.41 \\
UT1 & $G$ & $V$ (Vega) & 0.108 & 21.67 & 21.37 & 76\% & 18.89 & 18.76 & 18.64 \\
UT1 & $B$ & $g'$ (AB) & 0.173 & 21.66 & 21.49 & 85\% & 18.82 & 18.68 & 18.56 \\ \hline
UT2 & $L$ & $g'$ (AB) & 0.108 & 22.63 & 22.65 & 103\% & 19.82 & 19.56 & 19.37\\
UT3 & $L$ & $g'$ (AB) & 0.108 & 22.63 & 22.54 & 92\% & 19.71 & 19.45  & 19.26 \\
UT4 & $L$ & $g'$ (AB) & 0.108 & 22.63 & 22.45 & 85\% & 19.62 & 19.36 & 19.17 \\
\end{tabular}}
\end{table*}

The GOTO prototype was deployed towards the end of the second LIGO-Virgo Observing run (O2). The key goal was to ensure that the viability of both the design and implementation was confirmed so that a full facility could be built in time for the later observing runs. In the period between O2 and the start of the third observing period (O3) in early 2019, 
the prototype mount and unit telescopes were commissioned and tested, and upgrades were developed to improve the system performance and reliability. The telescope control, scheduling, image processing and source detection software was also developed during this period, to create a fully automated system from the point a transient alert is received to the potential sources appearing in the GOTO Marshall.

\subsubsection{Detectors}
\label{sec:detectors}

Each GOTO unit telescope is equipped with a 50 megapixel FLI MicroLine camera (see section~\ref{sec:hardware}). The physical properties of the detectors are given in Table~\ref{tab:specs}, and other parameters were measured prior to the cameras being shipped to La Palma for commissioning \citep[for details see][]{DyerThesis}. The gain, readout noise and fixed-pattern noise for each camera were measured using the photon transfer curve method \citep{janesick01}; each camera has a gain of between 0.53 and 0.63~e$^-$/ADU, with a typical readout noise of 12~e$^-$ and a fixed-pattern noise of 0.4 per cent of full-well capacity. By taking a series of long, dark exposures the dark current noise was measured to be less than 0.002~e$^-$/s for each camera. The cameras also each have a non-linearity of less than 0.2 per cent over their dynamic range, aside from when taking very short exposures or when close to saturation.

\subsubsection{Optics}
\label{sec:optics}

We measured the image quality of the GOTO-4 prototype using the full-width at half-maximum (FWHM) of all stellar sources across the field with airmass less than 1.2, using data from across March 2018. Under ideal observing conditions, the typical PSF at the centre of the frame was determined to have FWHM$\sim 2.5$ arcsec. Due to the inherent wide FoV, the PSF may show significant deviations (on average up to $\sim64$ per cent) between the centre of the frame and the edges. We found that the FWHMs in $L$, $R$, $G$ and $B$ bands are largely similar and found that the average FWHM values at the centre of the frame $\sim 2.5-3.0$ arcsec. 

The PSF performance was somewhat worse than expected ($1.8-2.5$ arcsec theoretical performance), in particular towards the field edges. However, we will see in the next sub-sections that it still allowed the prototype to deliver the necessary sensitivity and depth. Extensive tests were performed on the PSF behaviour and a number of issues were identified that contributed to this. Some optics and tube hardware upgrades were installed to mitigate these, and these issues informed the design of the next generation tubes to be used in the full facility (see \S \ref{sec:conclusions}). Key components in this were the stability of the primary mirror cell, the alignment of the corrector optics, and mount jitter.

We measured the vignetting across each instrument using the flat-field frames and find that the typical flux values deviate $\sim$10 per cent between the centre of the frame and the edges. The centre of the vignetting pattern is located approximately on the central pixels which suggests that the cameras were centred close to the line-of-sight of the optical axis. Additionally, we determined the amount of scattered light by analysing the large-scale deviations of flat field during dark and bright time. We found that the difference between the dark and bright conditions increase the overall background of the flat field by a factor of 2. The addition of cloths and baffling to the OTAs marked a significant improvement over the original design. Prior to this, the scattered light showed non-trivial structural gradients, which were subsequently removed by the additional baffling and allowed for more relaxed moon constraints.

\subsubsection{Sensitivity and Zero-point calibration}
\label{sec:sensitivity}

\begin{figure}
	\includegraphics[width=\columnwidth]{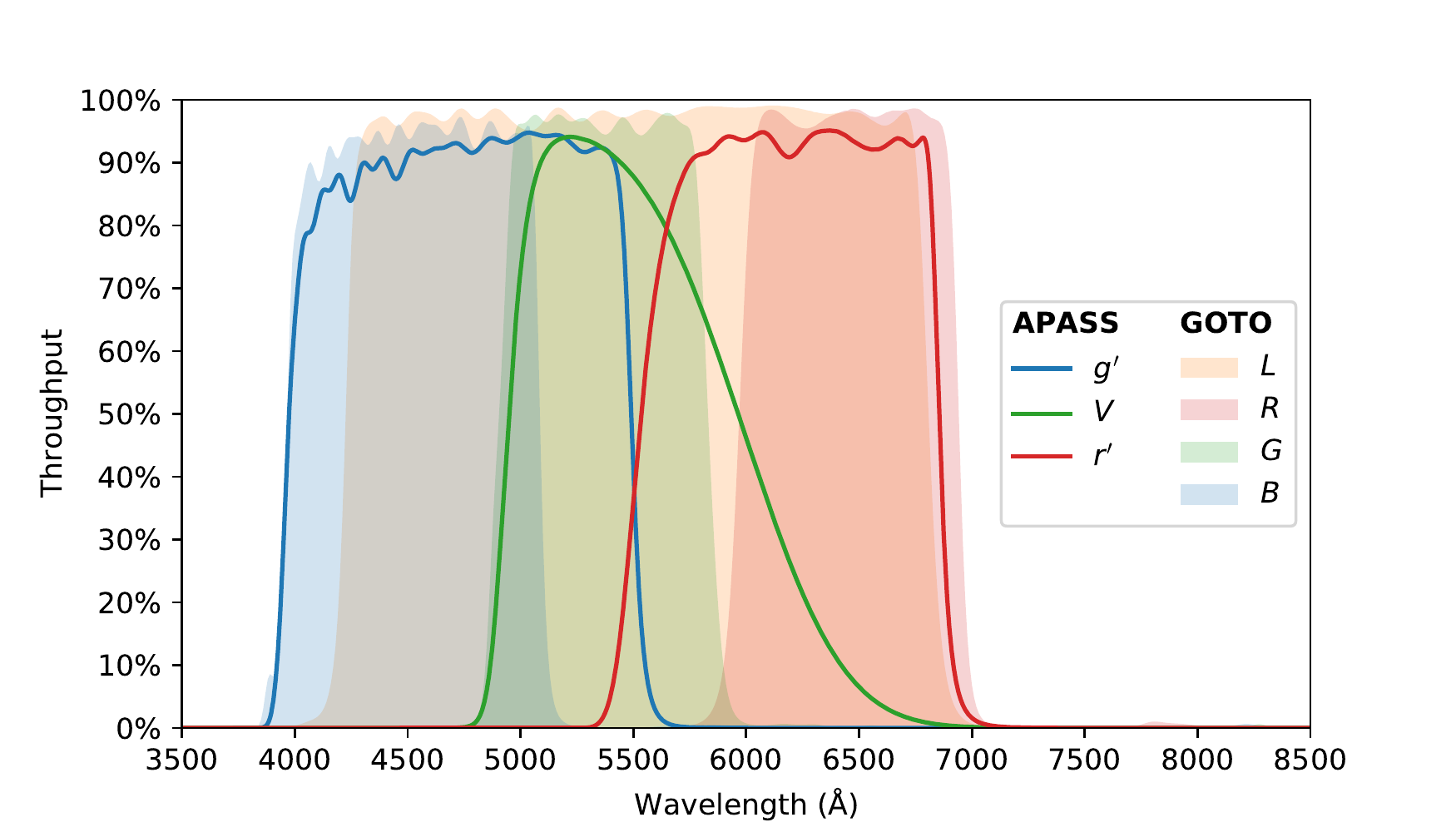}
    \caption{Bandpass comparison between the four Baader filters used by GOTO (filled areas) and the selected reference filters from the APASS survey (solid lines).}
    \label{fig:filters}
\end{figure}

\begin{figure}
	\includegraphics[width=\columnwidth]{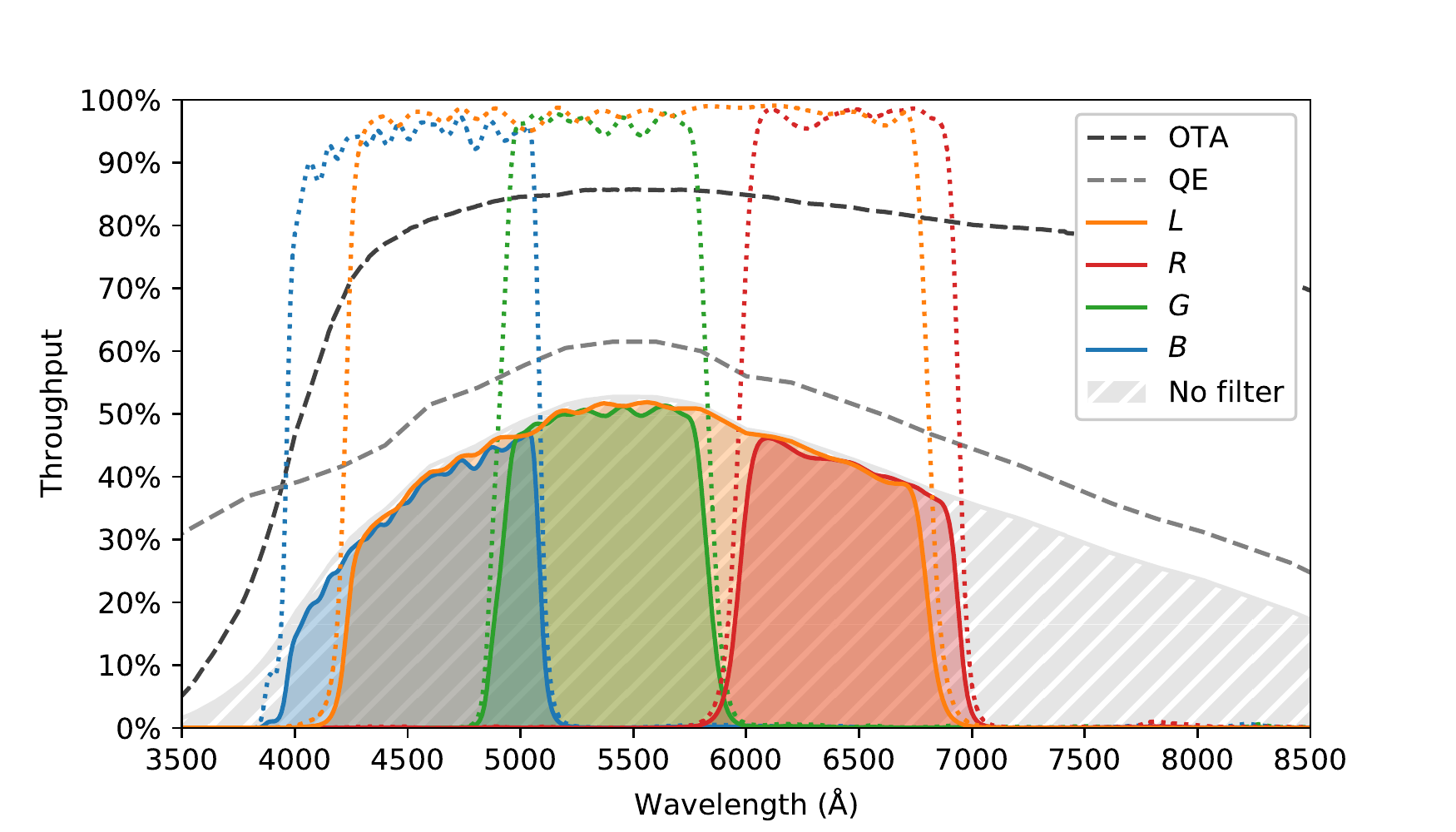}
    \caption{Throughput model for one of the GOTO-4 unit telescopes. The complete model (coloured areas) includes contributions from the OTA optics and CCD quantum efficiency (QE; dashed lines) and the bandpasses of the four Baader filters (dotted lines, from Fig.~\ref{fig:filters}). The grey hashed area shows the throughput of the system without a filter.}
    \label{fig:throughput}
\end{figure}

\begin{figure}
	\includegraphics[width=\columnwidth]{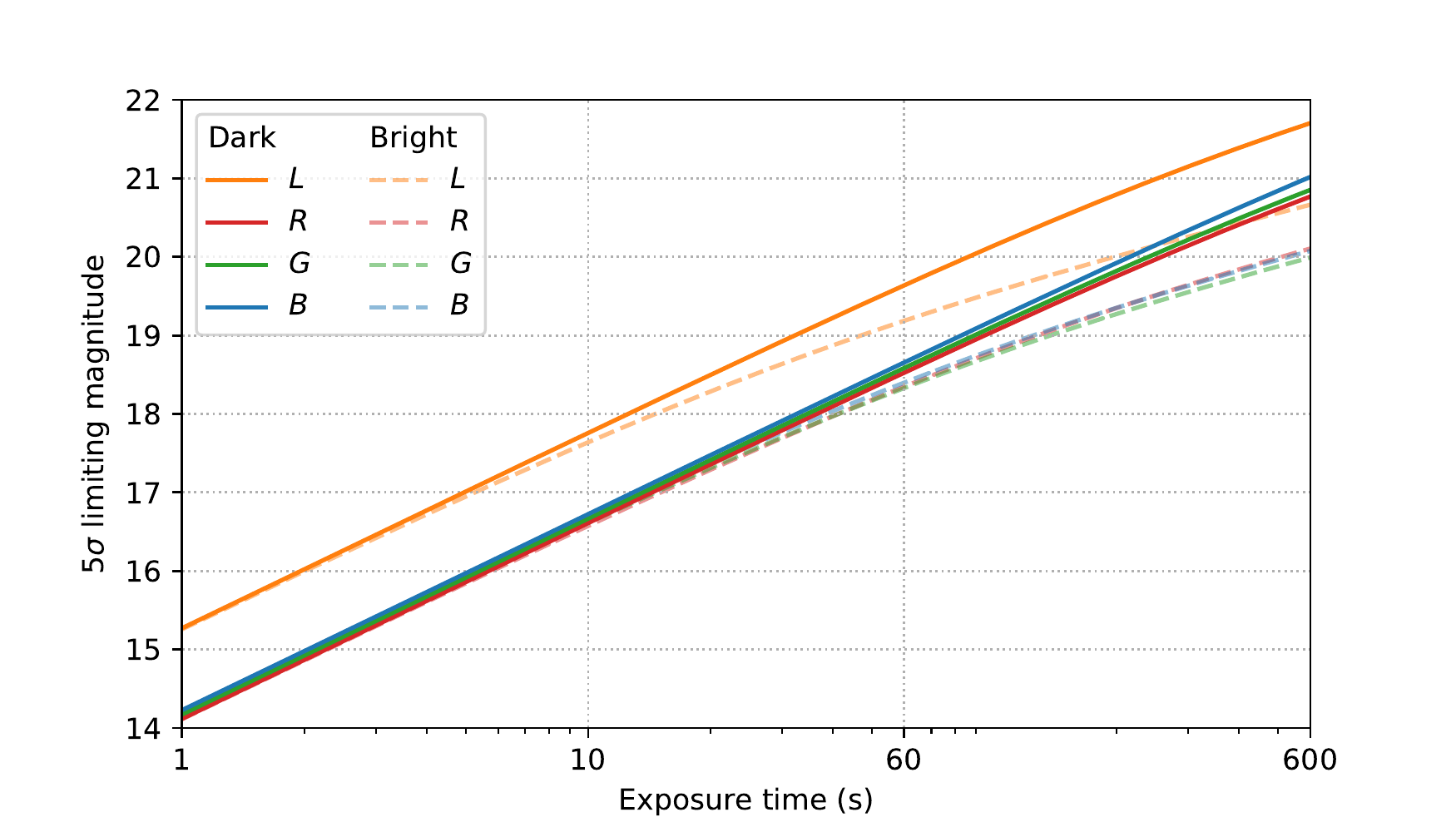}
    \caption{Calculated 5$\sigma$ limiting magnitudes for the GOTO-4 prototype as a function of exposure time, in the four Baader filters. Limits for dark and bright time are shown by solid and dashed lines respectively, and assume a target at airmass 1.0 and seeing of 1.5 arcsec.}
    \label{fig:lim_mag}
\end{figure}

The magnitude zeropoints were calibrated against the AAVSO Photometric All-Sky Survey (APASS) survey\footnote{\url{http://www.aavso.org/apass}}. The APASS survey is an all-sky photometric survey conducted in eight filters: Johnson $B$ and $V$ (in Vega magnitudes) and Sloan $u'$, $g'$, $r'$, $i'$, $z\_s$, and $Z$ (in AB magnitudes). Each GOTO frame was calibrated against the photometry from a set of referenced filters from the APASS survey. The first crossmatch is performed with {\sc catsHTM} using a cone search to identify the HDF5 APASS file. For each frame, all unsaturated ($L>14$) sources were spatially cross-matched to neutral colour ($-0.5 < g-r < 1$) APASS sources via a \texttt{KDSphere} cross-match. The reference filters were chosen based on the maximum integrated overlap area between the GOTO and the Johnson/Sloan filter response curves (Figure~\ref{fig:filters}): GOTO-$L$ is calibrated against APASS-Sloan-$g'$, $R$ against $r'$, $G$ against $V$, and $B$ against $g'$. As the $L$-band filter is broad, it essentially covers Sloan $g'$, $r'$ and Johnson $V$. However, since these zeropoints are to demonstrate headline performance of a prototype, we provide nominal zeropoints based solely against $g'$. For the characterisation of the final hardware, a more accurate prescription will be in place.

A throughput model of a GOTO unit telescope was constructed in order to determine the throughput of the system \citep{DyerThesis}\footnote{Data files are available from \url{https://github.com/GOTO-OBS/public_resources} and though the SVO Filter Profile Service (\url{http://svo2.cab.inta-csic.es/theory/fps/?gname=GOTO}).}. This model, shown in Fig.~\ref{fig:throughput}, includes the reflectivity of the primary and secondary mirrors, the transmission of the three lenses in the Wynne corrector and the glass window in front of the camera (collectively combined into the OTA throughput in Fig.~\ref{fig:throughput}), the QE of the CCD sensors, and the bandpass of each filter.

Using the \texttt{Astrolib PySynphot} package \citep{pysynphot}, theoretical zeropoints were calculated by passing the flux profile of a zero-magnitude star through the complete throughput model. These were $L_{ZP}$ = 22.63~mag (AB), $R_{ZP}$ = 21.33~mag (AB), $G_{ZP}$ = 21.67~mag (Vega) and $B_{ZP}$ = 21.66~mag (AB). Under typical observing conditions and during dark time, the airmass-corrected zeropoint magnitudes for a single UT (UT1) were observed to be $L_{ZP}$ = 22.47~mag (AB), $R_{ZP}$ = 21.27~mag (AB), $G_{ZP}$ = 21.37~mag (Vega), and $B_{ZP}$ = 21.49~mag (AB). For each UT, the airmass-corrected zeropoint magnitudes were found to be $L_{ZP}$ = 22.65, 22.54 and 22.45 for UT2, UT3 and UT4, respectively.

Based on the calibrated zeropoint magnitudes, the 5$\sigma$ limiting magnitudes that GOTO-4 was able to achieve are shown in Fig.~\ref{fig:lim_mag}. For a standard 60 second exposure a limiting magnitude of $L=19.8$ was predicted, which matches exactly the typical observed limits of $L=19.8$ during dark time and $L=19.56$ on average over all lunar phases. The modelled 5$\sigma$ limiting magnitudes are given in Table \ref{tab:performance} for all filters $LRGB$ for a single UT under dark, grey, and bright time and for all UTs for $L$-band under dark, grey, and bright time. The quoted performance is measured as $10^{(ZP - ZP_\mathrm{model})/2.5}$. Under all conditions, the calibrated zeropoints match reasonably well to theoretical expectations, despite the lower than expected performance characteristics of the PSF.

\subsubsection{Mount pointing \& tracking}
\label{sec:mount}

The pointing accuracy of GOTO was complicated by the array design, with each UT being affected by flexure in the mount, the boom-arm and the guidemounts holding each OTA (see section~\ref{sec:hardware}). The pointing accuracy is typically 2--5' but can be worse than 10' in declination, depending on the elevation. This, however, is still a small fraction of GOTO's large field of view, and future mount upgrades should reduce this further.

For similar reasons, the tracking could drift up to 1'/hour depending on the unit telescope. As GOTO typically only uses exposure times of 60 or 120 seconds, and only stays on each target for less than 5 minutes, this is rarely a major issue. Of more concern was the sensitivity to wind load as wind gusts can induce significant tracking errors. The prototype was particularly vulnerable to wind shake due to the exposed clamshell dome. Even under lower wind loads, the mount jitter contributed to the overall image PSF. The wormwheel design means that the motor torque is transferred via a belt and wormgear, which cannot be overly stiff.

\subsubsection{Sky coverage}
\label{sec:coverage}

\begin{figure*}
    \centering
    \includegraphics[width=0.9\linewidth]{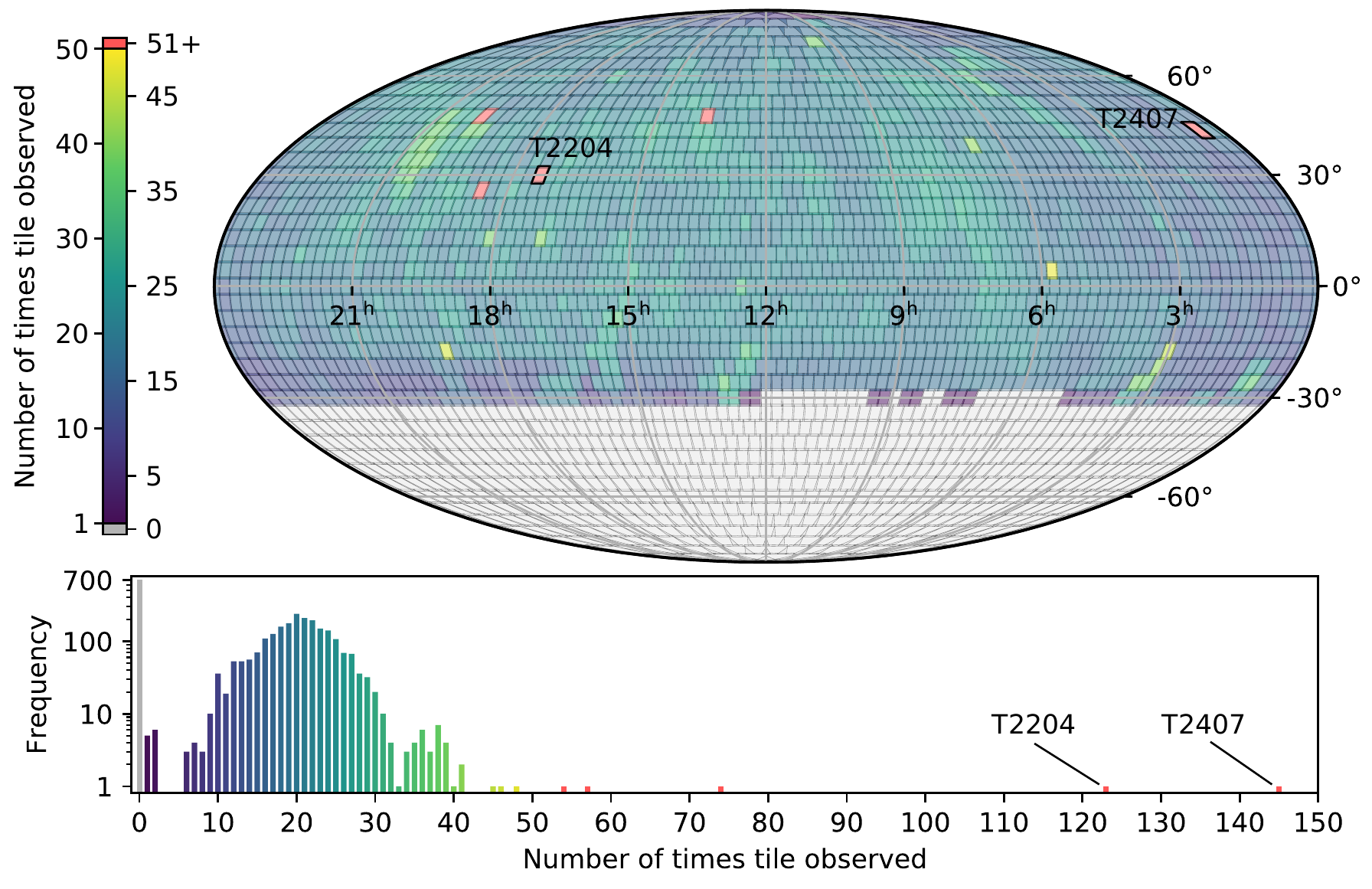}
    \caption{All on-grid observations taken by the GOTO-4 prototype between 21 February 2019 and 1 August 2020.
    The labeled tiles mark objects of particular interest during our commissioning observations; T2407 covers M31, and T2204 includes the GOTO2019hope field (see section~\ref{sec:coverage})}
    \label{fig:tile_counts}
\end{figure*}

The complete GOTO-4 prototype began to take regular observations in the evening of 21 February 2019, and covered the entire LIGO-Virgo O3 period from 1 April 2019 to its suspension on 23 March 2020. Afterwards, GOTO continued regularly observing until the morning of 1 August 2020, when the prototype was shut down in order to upgrade it to a full 8-UT array (see section~\ref{sec:conclusions}).

Between 21 February 2019 and 1 August 2020, GOTO observed at least one target on each of the 430 out of 527 nights (81.6 per cent), with the other nights in downtime due to bad weather, technical work and 53 days between 14 March and 6 May when the observatory was closed due to the COVID-19 pandemic. During this time GOTO observed 45,315 individual pointings, of which the vast majority (45,299 or 99.96 per cent) were aligned to the all-sky grid (see \ref{sec:scheduling}). The coverage of the on-grid pointings are shown in Figure~\ref{fig:tile_counts}. Of the 2913 tiles in the all-sky grid, 2207 (75.8 per cent) were observed at least once, with the remaining 706 being below the horizon visible from La Palma. The median number of observations per tile was 20. Two tiles were observed more than 100 times and are highlighted in Figure~\ref{fig:tile_counts}: T2407 contains M31 (00:42:44.3, +41:16:09) and was observed on 145 occasions, while T2204 contained GOTO2019hope (SN 2019pjv) (17:14:34.817, +28:07:26.26; see section \ref{sec:serendipity}) and was observed 123 times.

\subsubsection{Response time}
\label{sec:response}

The scheduling system described in section~\ref{sec:scheduling} allows GOTO to respond rapidly to transient alerts, and for events that occur during a clear night on La Palma GOTO can be observed within minutes. Of the eight gravitational-wave alerts to occur during clear nights in the commissioning period, observations of all but one began within 60 seconds after the alert was received \citep{DyerThesis}. The shortest time between an alert being received by the G\nobreakdash-TeCS sentinel and the exposures beginning was 28 seconds (for gravitational-wave event S190521g), and most of this delay was the unavoidable time spent slewing the telescope to the new target (see section~\ref{sec:gw}). Similar response times were recorded during GOTO's follow-up to GRB alerts. Under clear conditions and without any extraneous observational issues garnering delay, the shortest times between receiving the GRB alert and starting the exposures were 55 seconds (for Swift trigger 959431) and 2.3 minutes (for \emph{Fermi} GBM trigger 573604668) \citep{Mong2021}.

Once images are taken they are automatically transferred to the Warwick data centre and processed as described in section~\ref{sec:pipeline}. The typical latencies for the data transfer from La Palma to the Warwick data centre are $\sim$10 seconds. Single frames are processed within $\sim$3--5 minutes and within $\sim$10--12 minutes of the final exposure for coadding and stacking sets of science frames. The mean time from the mid-point of the exposure to a candidate entry being uploaded to the GOTO Marshall was 30 minutes, across all sources detected in 2020. This value excludes any delays of more than 2 hours, which are more likely due to network down-time or other disruption; without excluding those cases the mean delay was 47 minutes. The aim of future pipeline development is to reduce this delay to 10--15 minutes, which includes improvements to the latencies and efficiencies for database ingestion.

\subsubsection{Photometric and astrometric accuracy}
\label{sec:accuracy}

Long-term stability and accuracy of the photometric and astrometric measurements is key for high-quality data products.
An assessment of the photometric and astrometric accuracy has been detailed in \citet{Mullaney2021} and \citet{Makrygianni2021} in the context of exploring the compatibility of next-generation real-time pipelines, i.e. the LSST stack, on GOTO-4 data. The observations were obtained by GOTO during regular survey mode between 24 February 2019 and 31 July 2019 and covers the region between $02\mathrm{h}< \alpha < 20\mathrm{h}$ and $-20^{\circ} < \delta < +90^{\circ}$, specifically avoiding the densest regions of the Galactic plane.

The LSST-stack measured astrometry and photometry was compared to matched sources from PanSTARRS DR1, and it was found that the measured source positions were accurate to $0.27\pm0.20$ arcsec, and the $L$-band photometry was accurate to $\sim$50\,mmag at $L\sim 16$\,mag and $\sim$200\,mmag at $L\sim 18$\,mag. These values are favourably comparable to those obtained using {\sc GOTOphoto} \citep{Mullaney2021}.

Repeatability tests were also conducted on the tiles with the greatest frequency of visits. It was found that the photometric precision is typically within $1-2$\,mmag for sources brighter than $L\sim 16$\,mag, within $\sim 3-6$\,mmag for sources between $16>L>18$\,mag and within 0.2\,mag RMS of the Pan-STARRS photometry for sources fainter than $L< 18$\,mag \citep{Makrygianni2021}.

Further improvements are expected as we transition to a new data flow with more robust photometric calibrations and source flux determinations.

\section{Example science opportunities}
\label{sec:results}

In this section, we present some example science opportunities with results obtained during the commissioning phase of the GOTO-4 prototype on La Palma. 

\subsection{Gravitational-wave triggers}
\label{sec:gw}

\begin{figure*}
    \centering
    \includegraphics[width=0.8\linewidth]{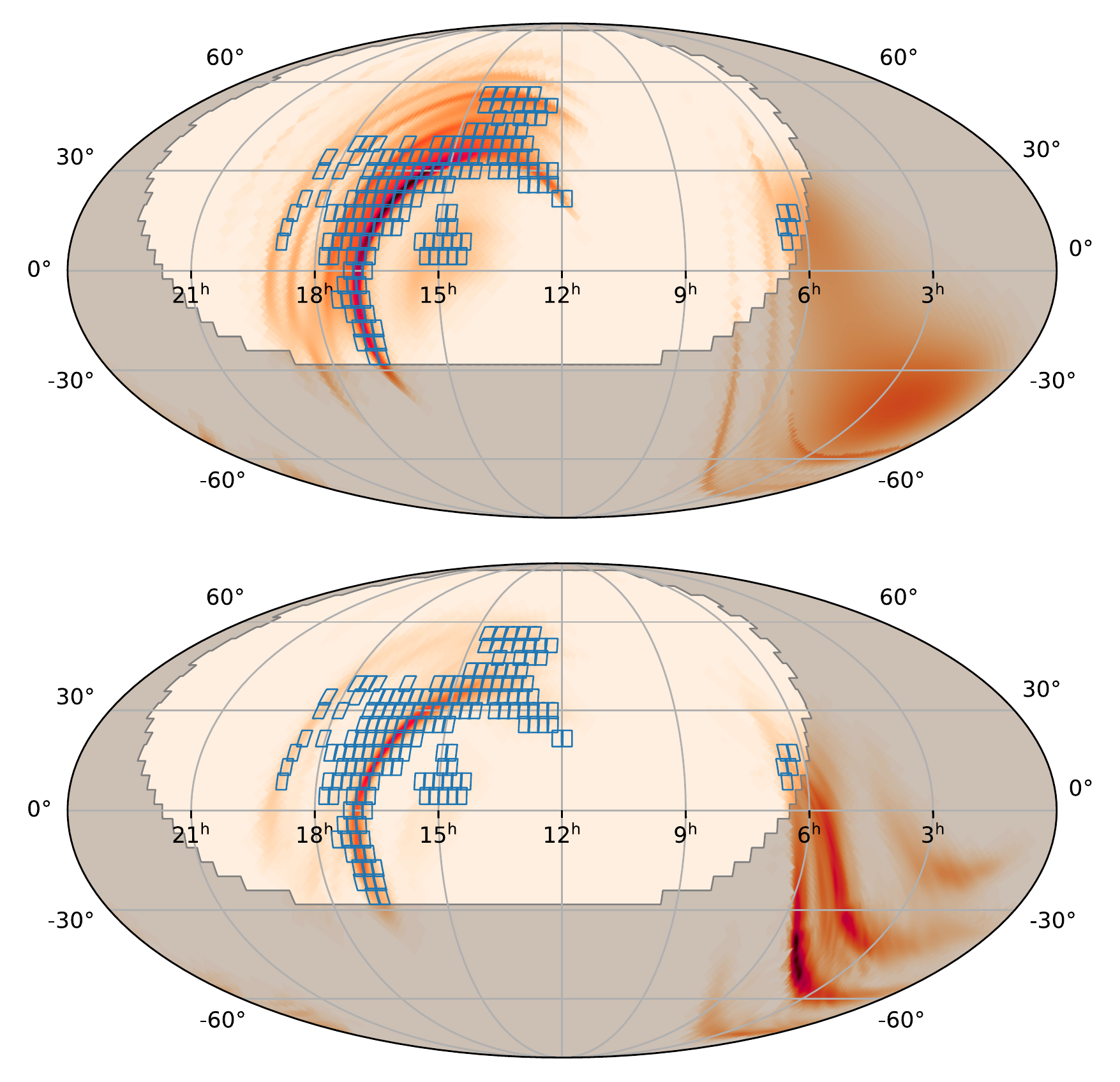}
    \caption{GOTO observations of GW190425 \citep{GW190425}, shown on the initial \textsc{bayestar} probability map (top) and final \textsc{lalinference} map (bottom). Blue squares represent individual pointings (tiles), and the orange shading shows probability density. Much of the probability initially resided near the well-covered northern crescent, but later shifted to the southern region during LVC re-analysis after the first observing night. The grey shaded areas were not observable from La Palma in the first three days after the event detection.}
    \label{fig:S190425z}
\end{figure*}

While the prototype GOTO-4 instrument was undergoing commissioning, the prioritisation was on targeting every GW trigger (regardless of source type) and the creation of a set of good-quality reference stacks for candidate counterpart identification. During the first half of the LVC O3 observing run, (O3a; April -- September 2019), the prototype GOTO-4 followed up 32 LVC GW triggers \citep[including 3 retractions; see][for a full summary]{Gompertz2020}. 
As noted in \S \ref{sec:response}, GOTO-4 can be on target within less than a minute from alert. The GW alert response time varied between $28$~seconds and $29.8$~hours, with an average of $8.79$ hours.
This large latency in the response time is mainly attributed to observational constraints, including the delay between the GW alert and the sky area becoming accessible from La Palma and weather conditions at the site.

In addition to rapid response capacity, GOTO also provides a unique set of wide-field capabilities, even with just 4 unit telescopes. This was particularly evident during the follow-up to GW190425 \citep{GW190425}.
The LVC alert was distributed during the La Palma day roughly 42 minutes after the GW event. The initial \textsc{bayestar} classification \citep{singer14, Singer15} was a BNS merger at a distance of $155 \pm 45$~Mpc. 
The 90 per cent credible region covered 10,183 square degrees \citepalias{LVCGCNS190425z}, with $71.1$ per cent observable from La Palma. GOTO-4 began observations nearly half a day after trigger imaging $\sim$2,134 square degrees (or 29.6 per cent of the skymap) during the first night. Shortly thereafter, the LVC probability map was updated using \textsc{lalinference} \citep{Aasi13,Veitch15}. While the distance and classification were largely unchanged, the new 90 per cent credible region was smaller \citepalias[down to 7,461 square degrees;][]{ligo2019ligo}, with much of the probability shifted to the unobservable southern sky (Figure~\ref{fig:S190425z}). GOTO-4 continued to observe the remaining $38.1$ per cent over the next two nights. Over the three-night campaign, GOTO-4 imaged 2,667 square degrees which included 37 per cent of the initial and 22 per cent of the final skymap. Although no counterpart was discovered, GOTO-4 was able to constrain the non-detection of an AT2017gfo-like kilonova out to 227~Mpc, or 6 per cent exclusion of the total volume of the LVC probability map \citep{Gompertz2020}.

Over the course of O3a, a mean of 732 square degrees were tiled per campaign, up to a maximum of 2,667 square degrees. GOTO-4 covered up to 94.4 per cent of the total LVC localisation probability, or 99.1 per cent of the observable probability. Of particular note is the inclusion of GOTO's data as part of an aggregate analysis of the follow-up to GW190814 \citep{GW190814}, the first potential neutron star -- black hole merger detected in GWs \citep{ENGRAVE20}, though it is now thought to be more likely a binary black hole merger \citep[][]{GW190814}. Given that the full GOTO facility will feature 8$\times$ the number of telescopes compared to the prototype, observations of GW sources will be a key strength of the facility.

\subsection{Gamma-ray bursts}
\label{sec:grbs}

In the absence of any prioritised GW trigger to follow-up, GOTO also participated in rapid follow-up of gamma-ray burst (GRB) triggers. Between 26 February 2019 and 07 June 2020, GOTO-4 observed 77 \emph{Fermi}-GBM and 29 \emph{Swift}-BAT burst alerts. 
GRBs were observed on a case-by-case basis to test different strategies and features of the observatory, and as such do not constitute a representative sample. However, taken as a group it can provide insight into the impact that GOTO can make in the explosive transients field. Further details on the overall performance of the GOTO-4 follow-up of GRB triggers can be found in \cite{Mong2021}.

During this time frame GOTO-4 detected four optical GRB counterparts, including the counterpart to GRB\,190202A which was detected at $L\sim 19$\,mag at $t\sim$ 2.2\,h after the trigger time \citep{GRB190202A} and the counterpart to GRB\,180914B detected $t\sim$ 2.15\,days post-trigger at $L\sim 20$ mag \citep{GRB180914B}. The observation response times for all GRBs ranged from 55\,s -- 69.3\,h after the GCN had been received by the G\nobreakdash-TeCS sentinel. Although a number of factors can determine the latency, observational constraints such as sky location and source rise time are the leading contributors to the measured latency rather than any significant instrumental delays.

Another notable example of GOTO's niche in this field was the response to GRB\,171205A.The observed photometric data points \citep{GRB171205A} complemented the other multiwavelength datasets which altogether describes a GRB that shows compelling evidence for the emergence of a cocoon \citep{Izzo2019}.

\subsection{Accreting Binaries}
\label{sec:acc_bin}
\begin{figure}
    \centering
    \includegraphics[width=\columnwidth]{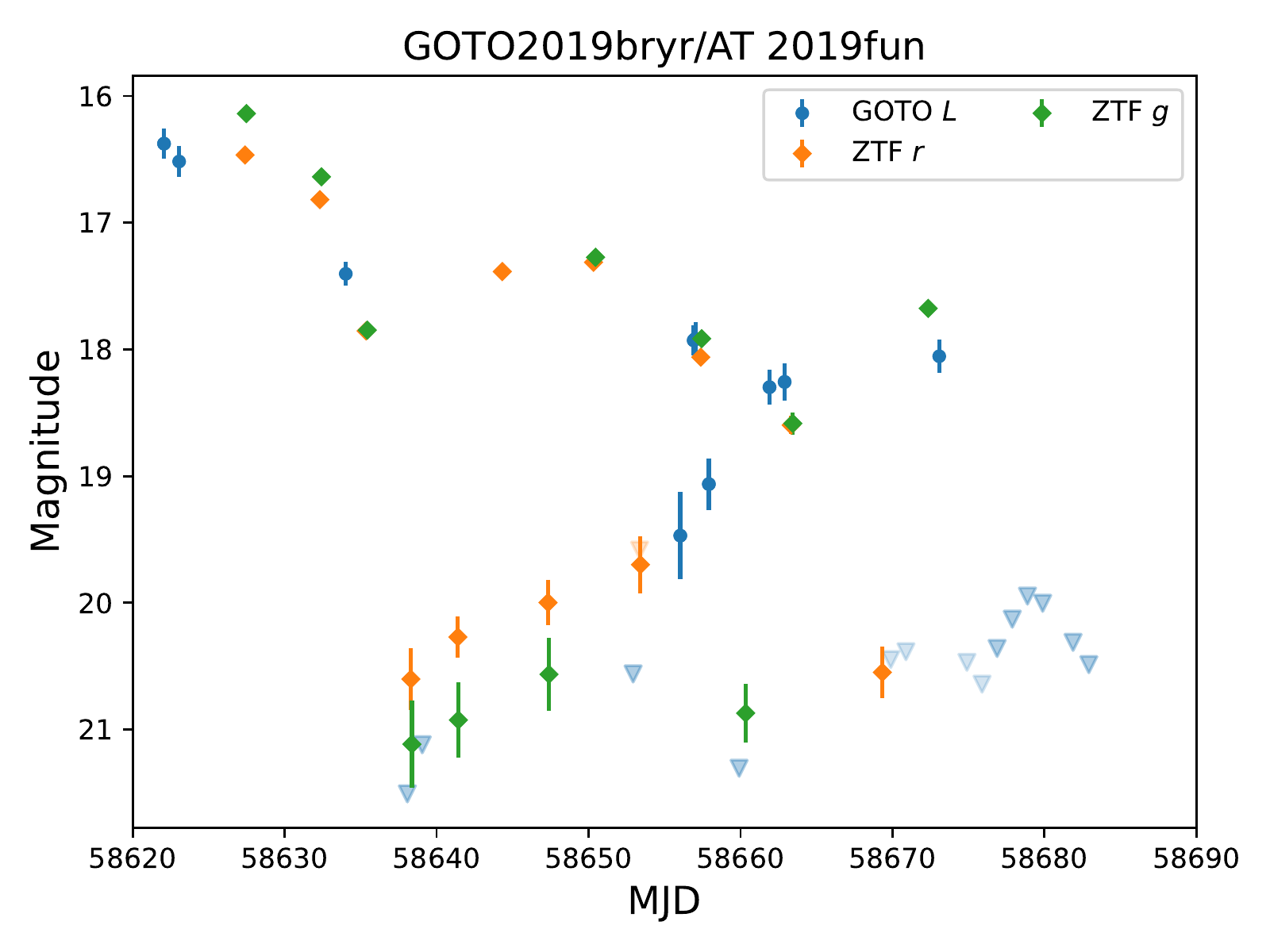}
    \caption{Light curve of the source GOTO2019bryr, also known as AT2019fun. This is a newly discovered CV, first detected by GOTO. Upper limits are marked with triangles. Contributions to the light curve from ZTF (r and g bands from Lasair) are also included for completeness.}
    \label{fig:GOTO2019bryr}
\end{figure}

Accreting compact binaries are a well established class of highly variable objects. Cataclysmic variables (CVs) are stellar binaries in which a white dwarf (WD) accretes matter from a nearby donor star. Within the compact accreting binaries family, CVs are far more abundant than their more massive counterparts known as X-ray binaries (XRBs), which harbour either a neutron star or a black hole. 
CVs are split into many subtypes depending on their average accretion rate, the magnetic field strength of the WD, the composition of the companion star, or the general behaviour of their light curve. Monitoring of 8 AM CVn systems with GOTO-4 and with long-term historical data sets revealed that there are diverse behaviours of a subset of AM CVn and that even within subclasses they may not be a homogeneous group \citep{Duffy2021}.

A common feature of a CV light curve is an increase in their luminosity by several magnitudes within a few days as their accretion disc undergoes a thermal instability \citep{Osaki1974}. Figure \ref{fig:GOTO2019bryr} shows an example of one of these so-called dwarf novae outbursts for a newly discovered CV (GOTO2019bryr / AT2019fun) observed by GOTO-4, where we have combined the median-stacked GOTO $L$-band data with photometry from ZTF\footnote{\url{https://lasair.roe.ac.uk/object/ZTF19aaviqnb}}, via the \textit{Lasair}   broker \citep{Lasair2019}. This system underwent several rebrightening epochs, which highlighted the need to monitor the long term evolution of this kind of outburst. 
The all-sky survey mode of GOTO will also be ideal for discovering new XRBs which enter into outburst as well, albeit at a lower detection rate than for CVs due to population sizes.

GOTO will also excel at identifying low level variations in the light curves of accreting binaries, which are likely due to small changes in the accretion rates in these systems. It has already contributed to the confirmation of a change in the accretion rate of the magnetic cataclysmic variable FO Aquarii \citep{Kennedy2019}. 

\subsection{Transients, Variables, and Moving Objects}
\label{sec:other_transients}

\begin{figure}
	\includegraphics[width=\columnwidth]{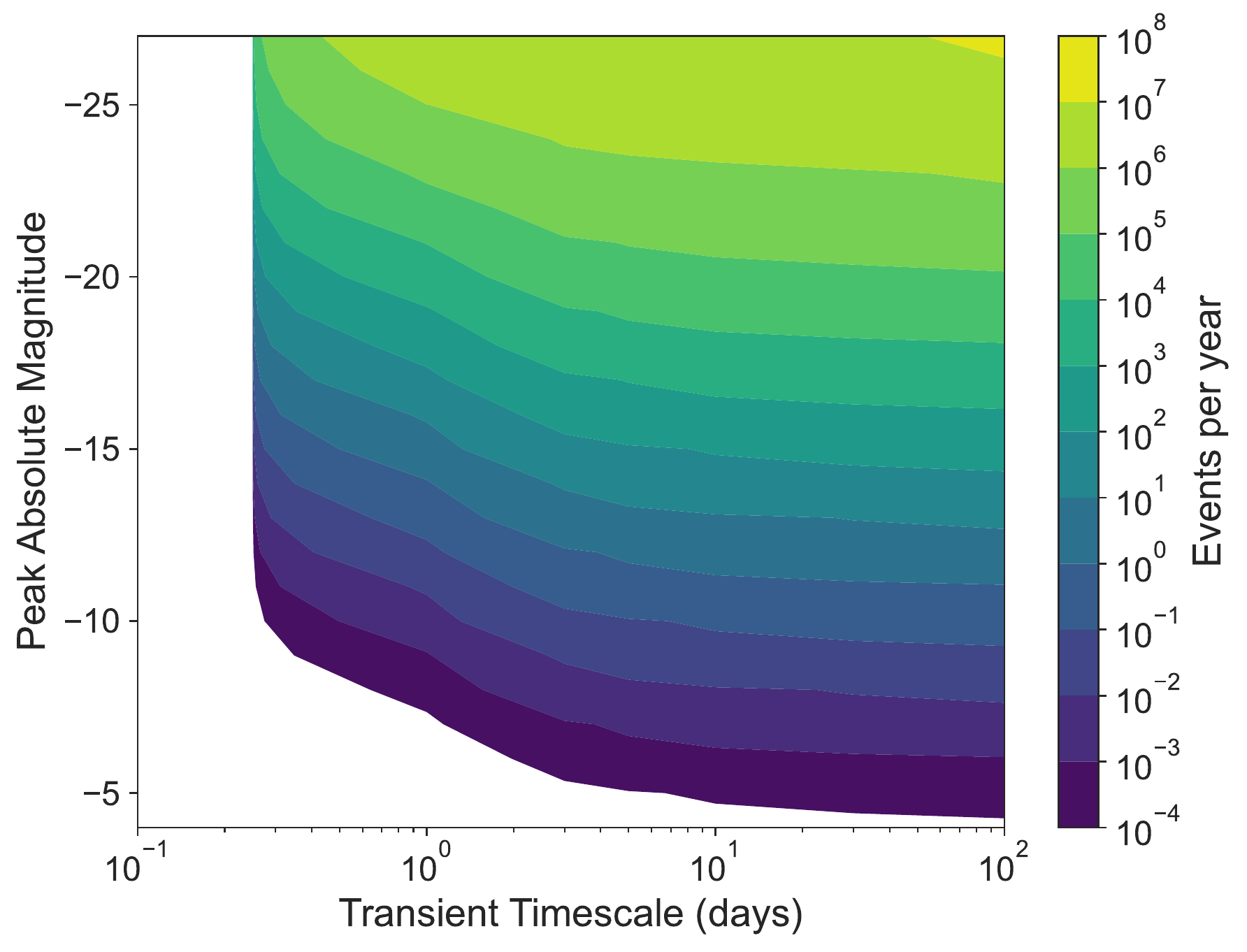}
    \caption{The event rate per year for a general transient that will be probed by a full GOTO-16 site as a function of peak absolute magnitude and decay timescales. For purposes of illustration we assume a 3 day cadence with a requirement of 2 consecutive detections. We grid over decay timescales ranging between 1 hour to 100 days per magnitude and over peak absolute magnitude between -4 to -27. We set a lower bound of events per year to $10^{-4}$\,yr$^{-1}$. This figure is has been created using the package described in \citep[][]{Bellm2016}.}
    \label{fig:transientrates}
\end{figure}

While many similarly-poised facilities undertake routine wide-area surveys, variances of cadence and depth determine the rate of expected numbers of variable and transient sources.
However, a general expectation of the rate of transients can be empirically estimated based on the site location, instrument hardware, survey sensitivity and cadence, among other considerations \citep[][; and adapted the described package\footnote{\url{https://github.com/ebellm/VolumetricSurveySpeed}} for our analysis]{Bellm2016}. Once GOTO moves into full operational mode, the entire available sky is expected to be covered every 3 days. To calculate the estimated rate of transient sources with GOTO-16, we assumed a 3-day cadence, that sources have been detected at least twice and have shown a decay rate of 1 magnitude over timescales between 1 hr and 100 days. The event rate per year covers the phase space as shown in Fig. \ref{fig:transientrates}. We gridded over possible combinations of peak absolute magnitudes and transient decay timescales which may represent generic transients and used the intrinsic rate of Type Ia supernova of $3.0\times 10^{-5}$\mpcyr as is the default setting of the package defined in \citep[][]{Bellm2016,LSSTScience2009}. The left edge boundary is an arbitrary cutoff bounded by a transient decay timescale of 1 hour to decline by 1 magnitude. The lower edge boundary is set by the lower bound of events per year for the illustration, or $10^{-4}$\,yr$^{-1}$.

There are a multitude of transient and variable astrophysical phenomena that are observable in the optical band. Transient events such as supernovae, flare stars, luminous red novae, dwarf novae outbursts, tidal disruption events, and kilonovae; and variable events such as, RR Lyrae, transits, eclipsing, rotating, and microlensing events, Active Galactic Nuclei (AGN) and BL-Lac objects, will all be routinely observed with GOTO in significant numbers. 
 
As a simple example to estimate the expected transient rates for a GOTO-16 system, we used the rates for typical Type Ia supernovae. We assumed 9 hours of observing per night down to a limiting magnitude of $L=19.8$ (a coverage of 11,520 square degrees per night). Given the peak absolute magnitude of a Type Ia SN of $M=-19$, a decay timescale of $\sim50$\,days, a volumetric rate of $3.0\times 10^{-5}$\,\mpcyr \citep{LSSTScience2009}, we find $\sim1596$ events per year.

Whilst not a core science goal, data from GOTO's general all-sky survey will uncover both known as well as unknown moving objects. The observing strategy of taking sets of 3-4 frames at each position will permit a direct search for objects moving on a short timescale. More rapidly moving objects will appear as apparent orphan transients and whilst being interlopers for some of the other science goals, there are excellent prospects for GOTO to contribute data concerning both new and poorly constrained moving objects. Initial effort on the detection of such moving objects made use of the {\sc CoLiTec} software \citep{Savanevych18}, which permits a semi-automatic search in parallel with the main pipeline. During commissioning, GOTO observed the near-Earth Apollo asteroid (3200) Phaethon. Quasi-simultaneous observations made with the Torino Polarimeter \citep[][]{Calern2012} mounted on the Omicron (west) telescope of the C2PU facility at the Calern observing station of the Observatoire de la C\^{o}te d'Azur obtained time-resolved imaging polarimetry and were used to probe the variation in surface mineralogy \citep[][]{borisov18}.

\subsection{Serendipitous Discoveries}
\label{sec:serendipity}

\begin{figure}
    \centering
    \includegraphics[width=\columnwidth]{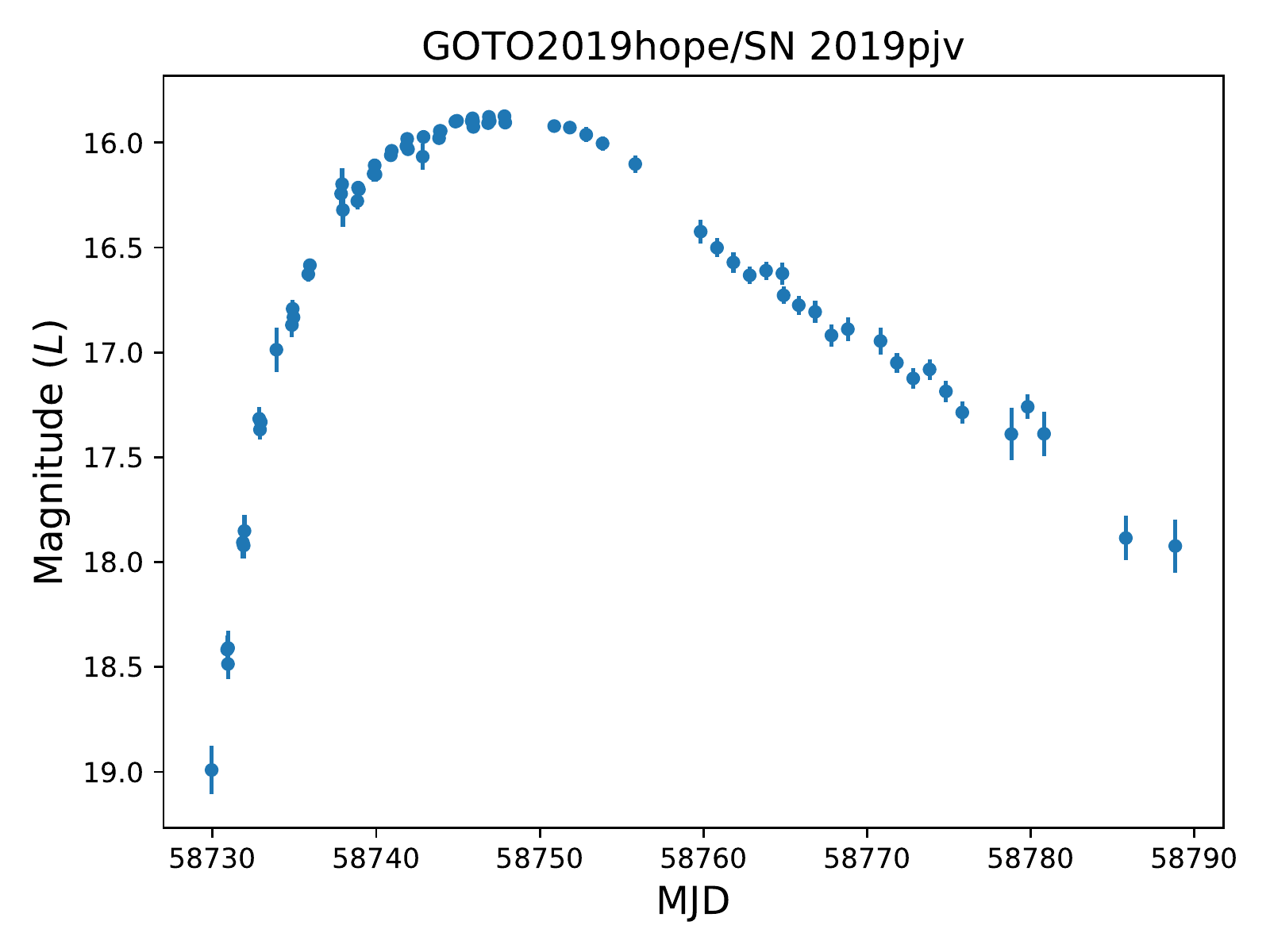}
    \caption{GOTO $L$-band light curve of GOTO2019hope/SN 2019pjv. This field was targeted nightly over the duration of the SN as a technical test for the difference image analysis and the transient ``realbogus" model.}
    \label{fig:GOTO2019hope}
\end{figure}

\begin{figure}
    \centering
    \includegraphics[width=\columnwidth]{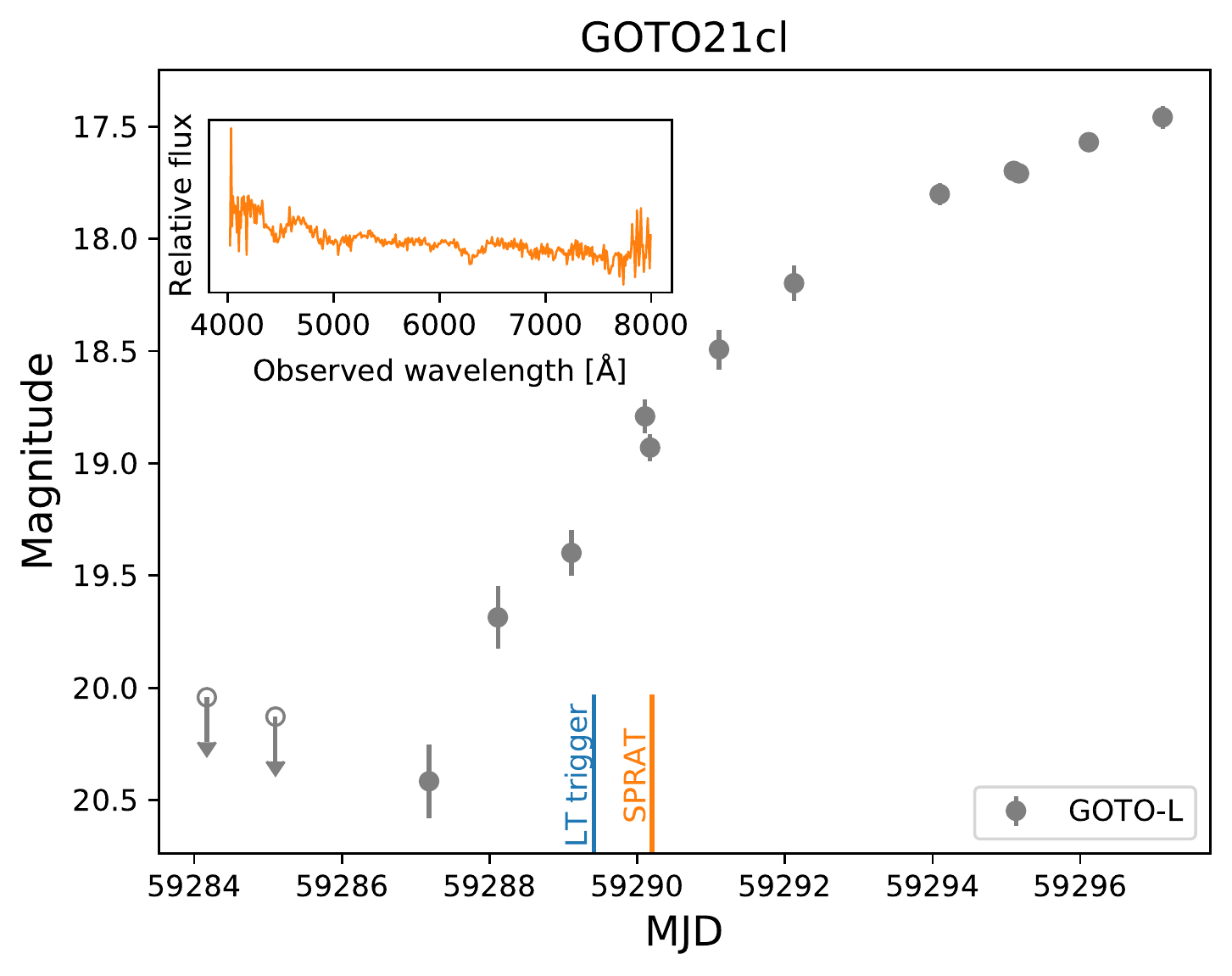}
    \caption{GOTO clear and $L$-band light curve of Type Ia SN GOTO21cl/2021fqb. The initial detection was picked up close to the noise limit of the detection image by the transient ``realbogus" model. The timescales for subsequent triggering of the Liverpool Telescope (LT) for follow-up and observations by LT's SPRAT instrument \citep{SPRAT2014} are shown.}
    \label{fig:GOTO21cl}
\end{figure}

During GOTO's long-term systematic survey campaigns or dedicated follow-up activities, it is also possible to make serendipitous discoveries. The novel and interesting serendipitous astronomical events must be filtered out amongst a variety of other sources that appear simultaneously in the images, such as transient impostors \citep[][]{Cowperthwaite2018,Pastorello2019,Almualla2021}; or as optical or instrumental contaminants, e.g. ghosts, cosmic rays, spurious noise. The identification of the novelty of the sources becomes particularly important when working under rapid identification timeframes for GW counterpart searches. While it is possible that identification of candidate transient events can be inferred through contextual information (such as whether there are previous non-detections of the source or if it is isolated, near a galaxy or within the galactic plane) often, and at first glance, they can appear to be legitimate transient events. While searching for the counterpart to the BNS GW candidate S190901ap \cite[]{gcn25606}, a serendipitous transient candidate was discovered. S190901ap was reported as a possible BNS out to an estimated distance $D_L$ = 241 $\pm$ 79 Mpc and was localized to an extremely large 14,753 square degrees (90 per cent) as reported by \textsc{bayestar} and \textsc{lalinference}. GOTO began observing the field 6.7 minutes post GW trigger time and followed the observing strategy for distant BNS sources \citep{dyer20,Gompertz2020}. Within the error region, a source of interest was identified with a detection magnitude of $L$=19.08 and a previous non-detection only 2.5 hours prior to the GW trigger time down to a limiting magnitude of $L \gtrsim 20$. The candidate was given the internal designation GOTO2019hope and the Astronomical Transient (AT) designation SN 2019pjv.

Within the field of view of the candidate were two possible host galaxies, MCG+05-41-001 and LEDA 1826843, separated by 46.92 and 64.31 arcsec and located at distances z=0.0227 ($D_L$=98.5 Mpc) and z=0.0707 ($D_L$=313.8 Mpc) respectively. Ultimately, spectroscopic follow-up of the source by GRAWITA on the Copernico 1.82m telescope, and later confirmed by the Nordic Optical Telescope (NOT), reported that the source best matches a Type Ia-91T like SN about one week before maximum light at a $z = 0.024$ \citep{GCN25661,GCN25665} and effectively ruled out the source as a transient associated with S190901ap.

While the source was ultimately deemed unrelated, it was a viable source for the GOTO-4 prototype to monitor long-term as a test field for photometric and astrometric accuracy monitoring, as well as testing of the "realbogus" model's stability (Section \ref{sec:sources}). Fig.~\ref{fig:GOTO2019hope} shows the rise, peak and decline of this Type Ia-91T SN.

Also with the serendipitous sources that GOTO will regularly observe, it is often meaningful to target early follow-up, in order to ascertain the relevance of these discoveries, and whether they warrant immediate follow-up. One example was GOTO21cl/SN 2021fqb as shown in Figure~\ref{fig:GOTO21cl}. This source was picked up during the routine patrol survey and is coincident (at 9.27 arcsec) with a luminous host galaxy at a redshift $z = 0.0490$ in the GLADE catalog. A spectrum of this source was taken with SPRAT \citep[Spectrograph for the Rapid Acquisition of Transients][]{SPRAT2014} which showed spectral features consistent with a young Type Ia supernova.

A key goal of the GOTO dataflow is to make the time delay between first detections and initial follow-up as short as possible by minimising dependencies on human vetting and flagging.

\section{Conclusions and future prospects}
\label{sec:conclusions}

The main purpose of the GOTO-4 prototype was to implement and further develop the concept of an array of medium-sized telescopes on shared mounts. The science focus is wide field time domain astronomy in the context of gravitational wave searches and other rapidly evolving objects. Whilst not without challenges, the performance of the GOTO-4 prototype instrument has clearly demonstrated the ability of the adopted design to meet the science goals.

\subsection{Lessons learned with the prototype}
Important lessons have been learned. First of all, the fast optics combined with a large sensor with small pixels places high demands on the UT implementation. The pixel scale is close to critical sampling to maximize field of view, and collimation and field correction needs to be tightly controlled.
Our prototype tubes highlighted the need for a stable primary mirror cell to control image quality stability at the edges of the field of view. Furthermore, scattered light can be an issue given the location of the corrector optics. For these reasons, the final GOTO UTs will feature closed carbon-fibre tubes with top end baffles as well as a more advanced primary mirror cell.
The second point was concerning the mount, which has to carry a heavy load as well as handle a big moment of inertia with tubes mounted far from the mount axes. Our prototype mount used a wormwheel design in an effort to be more robust against balancing. But in this design a small amount of mechanical slop in the various mechanisms that connect the mount motors to the axes meant a sensitivity to wind shake. The extensive footprint of the 8 tubes under a full load together with an open clamshell type enclosure further adds to this. Thus, for the final GOTO mount systems, a heavy-duty direct-drive system will be used to mitigate this.

The prototype instrument relied on the wide-band $L$ filter over the majority of the survey operations. This was a deliberate choice as it allowed for broadband response to transients without relying on color information. However, as is made apparent in the throughput model of Fig. \ref{fig:throughput}, there is a non-negligible sensitivity to the redder wavelengths. Future improvements to the instrument may include a custom wide-band red filter to enable coverage between $\sim$7000-8500\AA.

Finally, the prototype data-flow, {\sc GOTOphoto}, has been used successfully to benchmark and formulate the framework for the envisaged successor data-flow, which will need to have strong horizontal-scaling capabilities as the number of UTs used by GOTO increases. The new framework is in active development and in mid 2020 processed some stages of the data-flow in parallel. 

In addition to a requisite need for a more robust and scalable pipeline, so too is there for developing scalable transient identification algorithms, improvements to modelling wide-field PSFs for deconvolution and image subtraction, and automated image quality assessments for full-frame flagging. The new framework will address many of the early challenges that were uncovered during the prototype stage and will lead to technical advances in high-cadence wide-field optical image data processing.

\subsection{Vision for next phases}
\begin{figure}
	\includegraphics[width=\columnwidth]{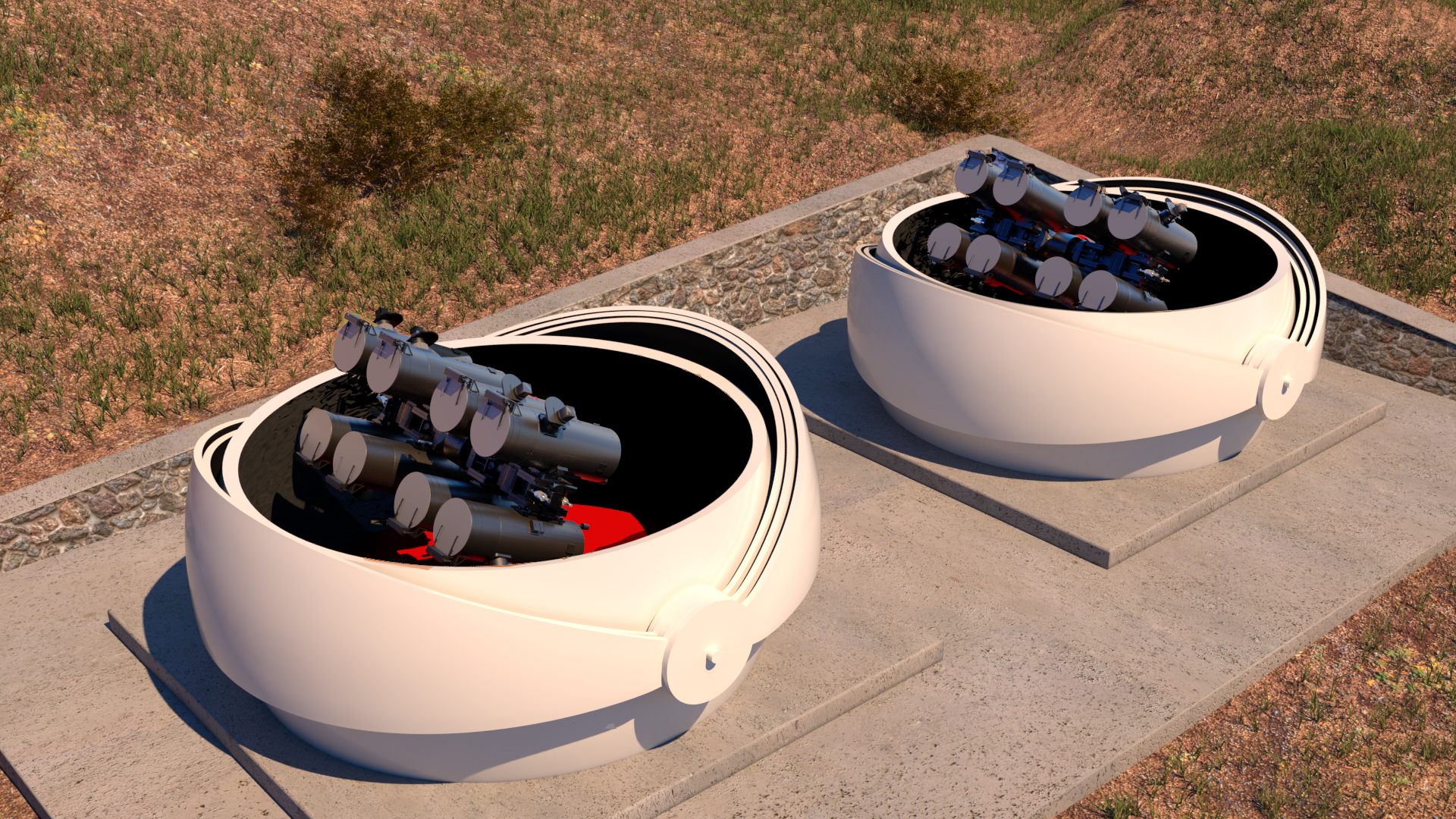}
    \caption{Visualisation of a full GOTO node site consisting of 16 UTs spread over 2 domes. The northern and southern nodes will contain identical sets of 2x8, providing a total of 32 UTs.}
    \label{fig:fullnode}
\end{figure}

With the UTs delivering the required headline performance metric, i.e. sufficient depth in a reasonably short exposure time, the true power then lies in deploying a significant number of telescopes across more than one location. In the GOTO design, the instantaneous footprint scales with the number of unit telescopes. The project is transitioning from prototype platform towards full deployment. At the La Palma site, the prototype equipment will be replaced by two new 8-telescope systems (Fig.~\ref{fig:fullnode}). These feature the revised tubes and mount noted above and will provide a collective field of view of $\approx75-80$~square degrees. With such a footprint, and a typical exposure set of several minutes per pointing, a good cadence can be achieved across the visible sky, a key driver for the project ($\sim 10,000$ square degrees per night). In parallel, a twin deployment is being developed at Siding Spring Observatory that will provide all-sky coverage. La Palma and Siding Spring are an ideal antipodal setup, covering all declinations whilst offering maximum complementarity. It is also of note that these two sites offer key longitudinal coverage compared to, for example Hawaii and Chile. The Siding Spring array will be identical to La Palma, also featuring two 8-telescope mount systems. 

This large expansion, amounting to an 8-fold increase in the number of telescopes deployed, will significantly boost the capabilities of GOTO compared to the prototype in both the monitoring and responsive modes. We previously considered an estimate for the expected rate of transients in survey mode for a dual 8-telescope system, or GOTO-16 in \S~\ref{sec:other_transients}. This is a full node at a given single site. With the addition of the second site at Siding Spring Observatory, the survey can be extended to cover all-sky and the expected rate of recoverable events will scale $\sim$linearly with coverage. Each site can cover about a quarter of the whole sky each night, and combined allows for an all-sky cadence of 2-3 days covering both N and S hemispheres. A considerable fraction of the sky is visible from both sites. However, there are also other modes available given the array approach of GOTO. It is also possible to point multiple mount systems at the same patch of sky to go deeper and reach events further in volume scaling as $\sim 1/r^2$. Alternatively, instead of maximising sky coverage or depth, the array could be split into groups that observe the same part of sky in different filters simultaneously. Colour information is particularly revealing for sources which show colour evolution or where colour information can reveal underlying properties of surrounding ejecta material -- as may be expected with kilonovae \citep[][]{Metzger2014BlueRed}.

GOTO is well situated in terms of a cost-effective, wide-field, scalable, and adaptable optical observatory. The volumetric survey speed \citep{Bellm2016} of a single site GOTO-16 can be estimated as $\sim 2\times 10^7$ Mpc$^3$/hr down to a limiting magnitude of $L=19.8$, comparable to that of other instruments, such as ATLAS and Pan-STARRS. For the full-scale GOTO instrument, the survey speed shows marked improvements, up to $\sim 10^8$ Mpc$^3$/hr for unique pointing strategy and $\sim 2\times 10^8$ Mpc$^3$/hr if overlapping alignment of tiles are observed. Similar metrics to the volumetric survey speed, such as the \emph{grasp} \citep[][]{Ofek2020}, can be used to showcase the unique niche that GOTO will fill in the current array of operable instruments. Based on estimates of information grasp, multiples of small and scalable telescope systems such as GOTO and ATLAS can be greater than 3 times more cost-effective compared to other survey telescopes with single unit systems.

Turning now to the responsive mode performance, we illustrate the impact on the core science area of EM counterpart searches coincident with GW triggers. A key improvement thanks to the dual anti-podal sites is an effective doubling of the duty cycle for any given GW event localisation region. This will significantly reduce the overall latency and roughly  double the number of recoverable events detected within, say, the first 12 hours. The single site nature of the GOTO-4 prototype was the main limiting factor setting the mean delay to first observation during the O3 run \citep[][]{Gompertz2020}.
The responsive mode searches will also profit directly from the increase in survey grasp. This will allow more search area to be covered more quickly. If we simply scale from the O3a sample in \citet[][]{Gompertz2020}, this would double the mean coverage to $\sim$1500 square degrees per campaign, or $\gtrsim$90\% of the O3a LVC probability skymaps and offer a reduction in the average response time delay of $\sim$4.5 hours.
A final key step is the continued evolution of the GW localisation performance, evolving to significantly better localisations as the global networks develop, and thus smaller areas to search over. The increase in survey grasp can then be used to provide denser and deeper coverage of these search areas. For events that are only accessible from a single site, this would be a fourfold increase in grasp, whereas events accessibly from both sites could receive the full factor of 8. Thus smaller areas and more telescopes combine to offer a significant opportunity to boost the typical depth achieved in search pointings.
Co-pointing multiple mount systems can be seen as essentially increasing the effective exposure time per set, with a corresponding gain in limiting magnitude (Fig.~\ref{fig:lim_mag}). The reduction in search area comes on top of this and will allow multiple visits to be stacked for even greater depth. Although the localisation performance and evolution is complex \citep[][]{LIGORates,petrov2021datadriven}, gains of 1-2 mags compared to the depth achieved in a single set are to be expected. 
The best strategy will be event-dependent, in terms of the specific optimal balance between maximising probability covered, depth achieved and time delay since GW trigger.

Looking in general at the prospects for kilonova detections for wide-field instruments also highlights how facilities such as GOTO complement and extend our search capabilities. The diverse specifications of current and planned facilities can be directly assessed in terms of the capability of probing the kilonovae detectable volume. Several studies have addressed the serendipitous detectability with the LSST at VRO \citep[e.g.][]{Cowperthwaite2019LSST, Setzer2019,Scolnic2018}, for other wide-field instruments like GOTO and DECam \citep[][]{Rosswog2017,Chase2021} and using infrastructure like the ZTF REaltime Search and Triggering \citep[ZTFReST][]{AndreoniCoughlin2021}. While space-based instruments like the Roman Space Telescope \citep[formerly WFIRST;][]{Spergel2015} are distinctly poised to reach deep ($z\sim 1$) into the volume, terrestrial observatories like LSST, DECam and GOTO are fully capable of imaging out to $z\sim 0.1$ \citep{Chase2021}. 

A more detailed description of the final GOTO hardware together with their performance metrics will be provided in a future paper, following commissioning of the science-grade arrays.

\section*{Acknowledgements}

We thank the referee Eric Bellm for the constructive comments.
The Gravitational-wave Optical Transient Observer (GOTO) project acknowledges the support of the Monash-Warwick Alliance; University of Warwick; Monash University; University of Sheffield; University of Leicester; Armagh Observatory \& Planetarium; the National Astronomical Research Institute of Thailand (NARIT); Instituto de Astrof\'isica de Canarias (IAC); University of Portsmouth; University of Turku; University of Manchester and the UK Science and Technology Facilities Council (STFC, grant numbers ST/T007184/1, ST/T003103/1 and ST/T000406/1). DS, DK, KA and YLM acknowledge support from the the Australian Research Council Centre of Excellence for Gravitational Wave Discovery (OzGrav), through project number CE170100004. JDL acknowledges fruitful discussions with Eran Ofek, Sergey Koposov and Dave Young. JDL acknowledges support from a UK Research and Innovation Fellowship(MR/T020784/1). H.K. was funded by the Academy of Finland projects 324504 and 328898. R.P.B., M.R.K., and D.M.-S. acknowledge support from the ERC under the European Union's Horizon 2020 re-search and innovation programme (grant agreement No. 715051; Spiders). D.M.S. also acknowledges the Fondo Europeo de Desarrollo Regional (FEDER) and the Canary Islands government for the financial support received in the form of a grant with number PROID2020010104. JL, BG, AJL and KW have received funding from the European Research Council (ERC) under the European Union's Seventh Framework Programme (FP7-2007-2013) (Grant agreement No. 725246, PI Levan).

This research has made use of data and/or services provided by the International Astronomical Union's Minor Planet Center. This research has also made use of the following Python packages: Astropy \citep{astropy13,astropy18}, numpy \citep{numpy20}, scipy \citep{scipy20}, healpy \citep{healpy19}, Astrolib PySynphot \citep{pysynphot}, astroplan \citep{astroplan18}, pandas \citep{pandas10}, photutils \citep{photutils20}, and scikit-learn \citep{scikit-learn}.

\section*{Data availability}
{Data files covering the system throughput and some of the software packages are available via public github repositories under \url{https://github.com/GOTO-OBS/}. Prototype data was mainly used for testing and commissioning and a full release of all data is not foreseen. Some data products will be available as part of planned GOTO public data releases.}




\bibliographystyle{mnras}
\bibliography{refs} 

\bsp	
\label{lastpage}
\end{document}